\newcommand{\be}{\begin{equation}}
\newcommand{\ee}{\end{equation}}
\newcommand{\bea}{\begin{eqnarray}}
\newcommand{\eea}{\end{eqnarray}}
\newcommand{\beann}{\begin{eqnarray*}}
\newcommand{\eeann}{\end{eqnarray*}}
\newcommand{\beasn}{\begin{sneqnarray}}
\newcommand{\eeasn}{\end{sneqnarray}}
\newcommand{\ba}{\begin{array}}
\newcommand{\ea}{\end{array}}
\newcommand{\nn}{\nonumber}
\newcommand{\Appendix}[1]%
    {\renewcommand{\thesection}{Appendix~\Alph{section}:}%
     \section{#1}%
     \renewcommand{\thesection}{\Alph{section}} }

\catcode`@=11
\def\secteqno{\@addtoreset{equation}{section}%
\def\theequation{\thesection.\arabic{equation}}}
\def\endsecteqno{\def\theequation{\@ifundefined{chapter}%
{\arabic{equation}}{\thechapter.\arabic{equation}}}}
\newcounter{subequation}
\def\thesubequation{\alph{subequation}}
\def\sneqnarray{\stepcounter{equation}\let\@currentlabel=\theequation
\setcounter{subequation}{1}
\def\@eqnnum{{\rm (\theequation\thesubequation)}}
\global\@eqcnt\z@\tabskip\@centering\let\\=\@eqncr\let\@@eqncr=\@@sneqncr
$$\halign to \displaywidth\bgroup\@eqnsel\hskip\@centering
 $\displaystyle\tabskip\z@{##}$&\global\@eqcnt\@ne
 \hskip 2\arraycolsep \hfil${##}$\hfil
 &\global\@eqcnt\tw@ \hskip 2\arraycolsep
$\displaystyle\tabskip\z@{##}$\hfil
  \tabskip\@centering&\llap{##}\tabskip\z@\cr}
\def\endsneqnarray{\@@sneqncr\egroup $$\global\@ignoretrue}
\def\@@sneqncr{\let\@tempa\relax
   \ifcase\@eqcnt \def\@tempa{& & &}\or \def\@tempa{& &}
   \else \def\@tempa{&}\fi
     \@tempa \if@eqnsw\@eqnnum\stepcounter{subequation}\fi
     \global\@eqnswtrue\global\@eqcnt\z@\cr}
\def\nobiblabels{\def\@lbibitem[##1]##2{\@bibitem{##2}}}
\catcode`@=12

\def\a{\alpha}  \def\b{\beta} \def\g{\gamma} 
\def\d{\delta} \def\D{\Delta} \def\e{\epsilon}
 \def\th{\theta}  
 \def\l{\lambda}  \def\m{\mu} \def\n{\nu}
    \def\r{\rho}
\def\s{\sigma} \def\t{\tau}  
    
\def\o{\omega}  

\newcommand{\NP}[3]{{\sl Nucl. Phys.} {\bf #1} (19#2) {#3}}

\newcommand{\PRL}[3]{{\sl Phys. Rev. Lett.} {\bf #1} (19#2) {#3}}
\newcommand{\ZFP}[3]{{\sl Z. Physik} {\bf #1} (19#2) {#3}}
\newcommand{\PR}[3]{{\sl Phys. Rev.} {\bf #1} (19#2) {#3}}
\newcommand{\AP}[3]{{\sl Ann. Phys. (N.Y.)} {\bf #1} (19#2) {#3}}

\def\zb{\bar{z}}

\def\apo{a^+_0}  \def\apz{a^+_z}  \def\apzb{a^+_{\zb}}  \def\ap3{a^+_3}
\def\amo{a^-_0}  \def\amz{a^-_z}  \def\amzb{a^-_{\zb}}  \def\am3{a^-_3}
\def\bo{a^3_0}         \def\b3{a^3_3}

\def\apm{a^+_{\mu}}
\def\amm{a^-_{\mu}}
\def\bm{a^3_{\mu}}

\def\dpzz{d^+_{zz}}   \def\dpzbzb{d^+_{\zb\zb}}
\def\dmzz{d^-_{zz}}   \def\dmzbzb{d^-_{\zb\zb}}
\def\ezz{d^3_{zz}}    \def\ezbzb{d^3_{\zb\zb}}

\def\dpz3{d^+_{3z}}   \def\dpzb3{d^+_{3\zb}}
\def\dmz3{d^-_{3z}}   \def\dmzb3{d^-_{3\zb}}
\def\ez3{d^3_{3z}}    \def\ezb3{d^3_{3\zb}}

\def\dpmn{d^+_{pq}}
\def\dmmn{d^-_{pq}}
\def\emn{d^3_{pq}}

\def\mpp{m^{++}}
\def\mmm{m^{--}}
\def\mm3{m^{-3}}
\def\mp3{m^{+3}}
\def\mmp{m^{-+}}
\def\md3{m^{33}}

\def\hd{h^{\dagger}}

\def\UdidU{U^{\dagger}i\partial_{\mu}U}

\def\Sv{{\bf S}}
\def\Dv{{\bf D}}

\def\xv{{\bf x}}
\def\nv{{\bf n}}

\def\pa{\partial} \def\da{\dagger} 

\documentstyle[12pt]{article}

\secteqno
\begin{document}


\title{{\bf Effective Field Theory Approach \\to \\Ferromagnets and
        Antiferromagnets \\in \\ Crystalline Solids }}
\author{{\Large {\sl Jos\'e Mar\'{\i}a Rom\'an}
                \ and \ {\sl Joan Soto}}\\
        \small{\it{Departament d'Estructura i Constituents
               de la Mat\`eria}}\\
        \small{\it{and}}\\
        \small{\it{Institut de F\'\i sica d'Altes Energies}}\\
        \small{\it{Universitat de Barcelona}}\\
        \small{\it{Diagonal, 647}}\\
        \small{\it{E-08028 Barcelona, Catalonia, Spain.}}\\
        {\it e-mails:} \small{roman@ecm.ub.es, soto@ecm.ub.es} }
\date{\today}

\maketitle
\thispagestyle{empty}


\begin{abstract}
We present a systematic construction of effective lagrangians  for the
low energy and momentum  region of ferromagnetic and antiferromagnetic 
spin waves in crystalline solids. We fully exploit the spontaneous symmetry 
breaking pattern $SU(2)\rightarrow U(1)$, the fact that spin waves are its 
associated Goldstone modes, the crystallographic space group and time reversal
symmetries. We show how to include explicit $SU(2)$ breaking terms due to
spin-orbit and magnetic dipole interactions. The coupling to electromagnetic 
fields is also discussed in detail. For definiteness we work with the 
space group $R\bar 3 c$ and present our results to next to leading order.
\end{abstract}

\bigskip
PACS: 11.30.Qc, 12.39.Fe, 75.30.Ds, 75.30.Gw, 75.50.Ee

\vfill
\vbox{
\hfill{cond-mat/9709298}\null\par
\hfill{UB-ECM-PF 97/23}\null\par}

%


\section{Introduction}
\indent

Whenever we have a lagrangian (hamiltonian) describing an infinite number 
of degrees of freedom with a given symmetry group $G$ the ground state of 
which has a smaller symmetry group $H$, we say that we are in a situation 
of spontaneous symmetry breaking (SSB, long range diagonal order). If in 
addition the symmetry corresponds to a continuous group we have gapless 
excitations. This is a consequence of the Goldstone's theorem, which was 
first proven for relativistic quantum field theories in \cite{Goldstone}, 
and then extended to nonrelativistic (condensed matter) systems
\cite{68,Nielsen,Anderson}.
We shall call Goldstone modes to the lowest lying (gapless) excitations.
In particle physics these correspond to massless particles and are known 
as Goldstone bosons. For instance in strong interaction physics, pions and 
kaons are the approximate Goldstone bosons of the approximate flavor 
$G=SU(3)_{L}\otimes SU(3)_{R}$ chiral symmetry of the QCD lagrangian which 
breaks spontaneously down to its vector part $H=SU(3)_{V}$ \cite{Weinberg}.
In condensed matter (CM) Goldstone modes appear in a large variety of 
systems describing quite distinct physics. Among the popular ones, which 
we shall be concerned with in the rest of the paper, are the magnons or 
spin waves in ferromagnets and antiferromagnets. These correspond to a 
$G=SU(2)$ spin symmetry being broken down to $H=U(1)$. Two more examples 
are the sound waves in a superfluid which correspond to the spontaneous
breaking down of the $G=U(1)$ particle number conservation to $H=Z_1$, and
the phonons in a crystal which are the Goldstone modes of the $G=T_3$
continuous translational symmetry to the discrete group $H$ of primitive 
translations of the crystal \cite{LeutwylerNR}. At somewhat more
complicated level it is perhaps worth mentioning superfluid $^3 He$ 
where $G=SO(3)\otimes SO(3) \otimes U(1)$ breaks down to   
$H=U(1)\otimes U(1)$, which presents interesting analogies with the
spontaneous symmetry breaking pattern of the electroweak theory
\cite{Volovik2}.

If one probes a physical system in the SSB phase by external sources
(e.g. electromagnetic waves) with small energy and momentum at low
temperatures, the only relevant degrees of freedom are the Goldstone modes. 
There is a systematic way to write down an effective lagrangian 
for the Goldstone modes which does not depend on the details of
the microscopic dynamics but only on the symmetry 
breaking pattern. For relativistic theories this is known from the late 60's 
\cite{Coleman}, but it has only been used extensively to next to leading order
for the last ten years \cite{Gasser} (see \cite{LeutwylerCEL} for a review).
It has been pointed out that the same techniques can be applied to condensed 
matter systems \cite{LeutwylerNR}, and lagrangians to the lowest order have
been provided for the ferromagnetic and antiferromagnetic spin waves,
and phonons \cite{Leutwyler}. In fact for antiferromagnetic spin waves
the effective lagrangian at leading and next to leading order has already
been used in \cite{Chakravarty} and \cite{Hasenfratz} respectively. 
However, the important r\^ole of the chrystallographic space group has 
been somewhat neglected.
For instance, in ref.~\cite{LeutwylerNR} it is claimed that symmetries alone 
cannot distinguish the effective lagrangian of the ferromagnet from that of the
antiferromagnet, and extra dynamical inputs (like the vanishing of the total
magnetisation for the antiferromagnet) are required. We shall see that once 
the space group is taken into account this is no longer necessary. In 
ref.~\cite{Burgess} the distinction between 
the ferromagnetic and antiferromagnetic effective lagrangians is made by 
imposing time reversal over macroscopic scales.
We shall also see that this requirement is too strong: whereas for some 
space groups having an antiferromagnetic ground state implies time 
reversal invariance at macroscopic scales for others it does not.
In the next to leading order calculation of ref.~\cite{Hasenfratz} Poincar\'e 
invariance is assumed instead of a crystallographic space group. 
We shall see that the crystal symmetries become even more important at next to 
leading order, allowing for new invariant terms in addition to the 
Poincar\'e invariant ones.

The power of these techniques relies on the fact that,
once the SSB pattern and space-time symmetries have been identified,
one can make a {\em controlled} expansion for any 
observable in terms of the typical energy and momentum of the Goldstone 
modes over the typical energy and momentum of the first gapped excitation.
The lagrangian is thus organised in terms of time and space derivatives.
For a given precision we only have to take into account derivatives until 
a given order. Then the lagrangian is a function of a few unknown 
constants which may be obtained from experimental data and be used later on 
to predict further experimental results. These constants may also be 
obtained by a complementary calculation {\em ab initio} starting from a 
particular model. Many examples of these calculations exist in the
literature (see for instance \cite{Dyson,Haldane,Dombre,Klee}). They 
are usually hard, require some approximations (typically Hartree-Fock or 
mean-field), and, of course, the outcome depends on the particular model 
choosen for the microscopic dynamics. We would like to stress that the 
effective lagrangian for the Goldstones modes encodes the low energy and 
momentum physics of {\it any} microscopic model with the same symmetry 
breaking pattern and space group, and hence it provides a model independent
description of the physical system to be studied. Different microscopic 
models just give different values for the unkwown constants.    

Recently, we have shown that the effective lagrangian techniques allow for 
an efficient description of non-reciprocal (time-reversal violating) 
effects in antiferromagnets in the microwave region \cite{Soto}.
The contribution of the spin waves to various (non linear) electric 
and magnetic susceptibilities has been calculated, which requires the 
knowledgement of the effective lagrangian at higher orders as well as 
the introduction of explicit symmetry breaking terms and the coupling to 
the electromagnetic field. Such a calculation would be extremelly difficult 
to carry out {\it ab initio} from any realistic microscopic model. 

In view of the above, and in order to stimulate further non-trivial 
applications of the effective lagrangian techniques, we feel it is worth
to illustrate here how the effective lagrangian for the spin waves in the 
ferromagnetic and antiferromagnetic crystals can be built in a systematic 
way by solely exploiting the symmetry breaking pattern and the relevant 
space-time group. We present our lagrangians at next to leading order.
We will also show how the electromagnetic interactions can be included
in the effective lagrangian. We will restrict ourselves to the space group 
$R\bar 3 c$, whose crystallographic point group is $\bar3m$, in order to make 
easy the comparation with ref.~\cite{Soto}, but it will be clear at any stage 
how to proceed for any other crystallographic group. The paper is intended 
to be self-contained.

In order to simplify the notation we will take $\hbar=c=1$ which leads
to a relativistic notation. So $x_{i}=(t,\xv_i)$, where subindex $i$
represents a lattice position, subindices \mbox{$\m=0,1,2,3$}, where the
first one represents the time component.

We distribute the paper as follows. 
In section~2 we describe the basic fields and symmetries. In section 3 we 
explain how to systematically construct the effective lagrangian to a given 
order of space and time derivatives. In sections~4~and~5 we present the 
effective lagrangian to next-to-leading order for the fe\-rro\-magnet and
antiferromagnet respectively. In sections~6~and~7 we show how to include 
$SU(2)$ breaking and magnetic dipole interactions respectively. In
section~8 the coupling to electromagnetic fields is included. Section~9 
is devoted to a discussion of our results. In Appendix~A we discuss the 
effect of non-trivial primitive translations in antiferromagnets. Appendix~B 
contains technical details. In Appendix~C we show some particular features 
for the fundamental representation, $s=1/2$. Appendix~D contains a proof of 
the equivalence between our formulation and the so-called $O(3)$-sigma model 
as well as a brief comment on the different forms that a certain topological 
term can be found in the literature. In Appen\-dix~E we show how constant 
electric and magnetic fields modify the spin wave dispersion relation.



\section{Effective Fields and Crystal Symmetries}

\subsection{Internal symmetries}
\indent

When $G$ is a compact internal symmetry group, namely, a compact group
disentangled from the space-time symmetry group, a general analysis was 
provided in ref.~\cite{Coleman}. Although the construction of the 
effective lagrangian was carried out for relativistic theories, it readily 
applies to non-relativistic theories by just changing the Poincar\'e group 
by the relevant space-time group. The outcome of ref. \cite{Coleman} is
that the effective lagrangian for the Goldstone modes can always be written 
in terms of a matrix field $U(x)$ taking values in the coset space $G/H$, 
determined by the pattern of symmetry breaking $G \rightarrow H$ (recall 
that $H$ is the internal symmetry group of the ground state). We are 
searching for an effective lagrangian invariant under $SU(2)$. The
field $U(x)$ transforms non-linearly under $SU(2)$ as follows:
\be
U(x) \rightarrow gU(x)\hd(g,U).
\label{U_SU2_transf}
\ee
When $g \in H$, the unbroken subgroup, then $\hd = g^{\da}$ and $U(x)$
transforms linearly.

In order to have an intuitive picture of the above mathematical formulation, 
it is helpful to take the Heisenberg model as a microscopic model,
\be
H = \sum_{<i,j>} J_{ij} \Sv_i \Sv_j.
\label{Heisenberg}
\ee
Our discussion however is general and holds for more complicated models, 
like the Hubbard model, $t-J$ model, etc., with the only requirement that they
have an $SU(2)$ spin symmetry which breaks spontaneously down to $U(1)$ in 
the ground state. (\ref{Heisenberg}) can be written in the second quantisation
language in terms of the real space creation and annihilation operators, 
namely, $\psi^{\dagger}(x)$ and $\psi(x)$,
\be
H = \sum_{<i,j>} J_{ij}
            \left(\psi^{\dagger}(x_i) \Sv \psi(x_i)\right)
            \left(\psi^{\dagger}(x_j) \Sv \psi(x_j)\right).
\label{second_quantisation}
\ee

This hamiltonian realises the internal $SU(2)$ symmetry as follows:
\be
\psi(x_i)\longrightarrow g\psi(x_i)  \quad ,\quad g\in SU(2).
\label{psi_transf}
\ee
As a classical field theory, the ground state configurations $\psi_0(x_i)$ 
of (\ref{second_quantisation}) are those with a maximum spin in a given 
direction, say the third direction. For spin~$1/2$ we have
\mbox{$\psi_0^{\da}(x_i) = (1 \ 0)$} for all $i$ in the ferromagnet,
whereas in the antiferromagnet half of the $i$s would have the above 
configuration whereas the remaining half would have 
$\psi_0^{\da}(x_i) = (0 \ 1)$. The symmetry of the ground state 
configurations is clearly $U(1)=\langle e^{i\theta S^3} \rangle $. We can 
think of a classical configuration close to the ground state as 
$\psi(x_i)\sim \tilde U(x_i) \psi_0(x_i)$, $\tilde U(x_i)$ being slowly 
varying through the lattice. $\tilde U(x_i)\in SU(2)$ admits a unique 
decomposition $\tilde U(x_i)=U(x_i)h(\tilde U(x_i))$ where 
$U(x_i)\in SU(2)/U(1)$ and $h\in U(1)$. Since the ground state configuration 
is $U(1)$ invariant we can always write $\psi(x_i) \sim  U(x_i)\psi_0(x_i)$. 
Taking into account this relation, (\ref{psi_transf}) and the $U(1)$ 
invariance of the ground state configuration, the non-linear transformations 
(\ref{U_SU2_transf}) are justified ($h^{\da}(g,U)$ is the suitable factor 
that left multiplied by $gU$ gives an element of the coset).

If the magnetisation occurs in the third direction then
\be
U(x) = exp\left\{ {i \sqrt{2}\over f_{\pi}}
               \left[ \pi_{1}(x)S^1 + \pi_{2}(x)S^2 \right] \right\},
\label{spin_waves}
\ee
where $S^i$, are  generators of $SU(2)$ for any representation and 
$\pi_{i}(x)$ are the fields describing the spin waves ($f_{\pi}$
is a dimensionful factor).

\subsection{Space-time symmetries}
\indent

Now that we have our basic field and know what its transformations are
under the internal symmetry group, we have to find out how it transforms
under the space-time symmetries. The space group and time reversal must 
be respected by the dynamics. The space-time symmetry will be broken by 
the ground state configuration. In fact the microscopic hamiltonian presents 
the symmetry of the paramagnetic fase ${\cal S} \otimes T$, i.e., space group
and time reversal, which breaks down to the ground state symmetry given
by the magnetic space group \cite{Birss}. As in the case of the internal
symmetry we impose our effective lagrangian to be invariant under the
unbroken symmetry ${\cal S} \otimes T$. For definiteness we will be
concerned with the space group $R\bar 3 c$, whose crystallographic  
point group is $\bar3m$, together with time reversal. The $\bar3m$ group 
is generated by a rotation of $2\pi / 3$ around the $z$-axis ($C_{3z}^+$), 
the inversion ($I$) and a reflexion plane perpendicular to the $y$-axis 
($\s_y$). The time reversal symmetry just reverses the sign of time. If 
we introduce holomorphic coordinates $z=x+iy$ and $\bar z = x-iy$ these 
transformations read
\be
   \ba{rl}
   C^+_{3z}: &
      \left\{ \ba{ccc}
      z   &  \rightarrow  &  e^{i2\pi/3}z     \\
      \zb &  \rightarrow  &  e^{-i2\pi/3}\zb  \\
      x^3 &  \rightarrow  &  x^3
      \ea  \right.
   \\
    &   \\
   I:  &
      \left\{ \ba{ccc}
      z    &  \rightarrow  &  -z               \\
      \zb  &  \rightarrow  &  -\zb             \\
      x^3  &  \rightarrow  &  -x^3
      \ea \right.
   \\
    &    \\
   \s_y:  &
      \left\{ \ba{ccc}
      z    &  \rightarrow  &  \zb              \\
      \zb  &  \rightarrow  &  z                \\
      x^3  &  \rightarrow  &  x^3.
      \ea \right.
   \ea
\label{space_transf}
\ee

We will consider $U(x)$ in the continuum, which is always a good approximation 
for small momentum. Effects due to finite lattice spacing are encoded in 
higher space derivative terms. In the continuum approach only the crystal 
point group  and the primitive translations ($\t$) of the full space group 
are relevant.

The ground state configuration can be arranged ferromagnetically or
antiferromagnetically. This leads to different transformations of the
$U(x)$ field under the crystal point group and primitive translations.

If the ground state is ferromagnetic the local magnetisation points
to the same direction everywhere. This indicates that we have to assign
trivial transformation properties to $U(x)$ under both the point group 
and the primitive translations.
\bea
   \ba{rccl}
   C^+_{3z}: & U(x)  &  \rightarrow  &  g_3U(x)\hd_3 \\
   I:        & U(x)  &  \rightarrow  &  U(x)         \\
   \s_y:     & U(x)  &  \rightarrow  &  g_2U(x)\hd_2 \\
             & & & \\
   \tau:     & U(x)  &  \rightarrow  &  U(x),
   \ea 
\label{U_F_space_transf}
\eea
where $(g_i, h_i)$ are the non-linear $SU(2)$ transformations induced
by the rotations. Notice that only the rotational part of the 
roto-translational elements which may exist in the space group 
is relevant and this is already included in (\ref{U_F_space_transf}).
This is a general feature independent of the particular space group.

On the other hand, if we have an antiferromagnetic ground state the 
local magnetisation points to opposite directions depending on the 
point of the space where the magnetic ion is located. This
must be reflected in the transformation properties of $U(x)$. Those
depend in turn on how the magnetic ions are distributed in the
crystal. In order to make it definite, let us consider the $Cr_2O_3$
crystal, which enjoys the $R\bar 3 c$ space group with $\bar3m$ point 
symmetry group. The rhombohedrical unit cell contains
four $Cr$ atoms located along the $z$-axis which play the role of the
magnetic ions, and six oxygen atoms which play no role as far as 
spin is concerned \cite{LB}.
However, the presence of oxygen atoms is crucial for the absence of
primitive translations which map points with opposite magnetisations.
Hence all primitive translations must be implemented trivially.
The point group symmetries $C_{3z}^+$ and $\s_y$ map points with the 
same local magnetisation, while the inversion $I$ maps points with
opposite local magnetisations. This indicates that we may assign to 
$U(x)$ the following transformation properties under the $\bar3m$ group:
\be
   \ba{rccl}
   C^+_{3z}: & U(x)  &  \rightarrow  &  g_3U(x)\hd_3  \\
   I:        & U(x)  &  \rightarrow  &  U(x)C\hd_I    \\
   \s_y:     & U(x)  &  \rightarrow  &  g_2U(x)\hd_2  \\
             & & & \\
   \tau:     & U(x)  &  \rightarrow  &  U(x)
   \ea
\quad , \quad
C = e^{-i\pi S^2},
\label{U_AF_space_transf}
\ee
$h_{I}$ is a compensating $U(1)$ element that keeps the transformed 
field in the coset and $C$ turns a spin up into a spin down. 

Even though the $Cr_2O_3$ does not have primitive translations which map 
points with opposite local magnetisations there are antiferromagnetic 
materials which do have them. In that case non-trivial transformations should 
be assigned to $U(x)$. This is discussed in detail in the Appendix A.

Notice then, that in the antiferromagnetic case, the transformations of 
$U(x)$ are dictated not only by the space symmetry group but also by the
precise local magnetisation of the points related by the symmetry operation.

Let us finally discuss time-reversal symmetry. This symmetry is
spontaneously broken in both ferromagnetic and antiferromagnetic ground
states. Time reversal changes the sign of the spin. This transformation
is implemented on a wave function by
$\psi(x) \rightarrow C\psi^*(x)$ \cite{Pascual}, which translates for
$U(x)$ field in
\be
   \ba{rccl}
   T: & U(x)  &  \rightarrow  &  U(x)C\hd_t
   \ea
\quad , \quad
C = e^{-i\pi S^2},
\label{U_time_transf}
\ee
$h_{t}$ is again a compensating $U(1)$ element that keeps the
transformed field in the
coset.

It can be shown that when the space-time transformations associated to
the magnetic space group, the unbroken subgroup of the space-time
group, are considered the transformations of $U(x)$ are linear.



\section{Construction of the Effective Lagrangian}

\subsection{Building blocks: Internal transformations}
\indent

After having established the field transformations in the previous 
section, we shall proceed to construct the effective lagrangian in 
terms of $U(x)$ and its derivatives. Following \cite{Coleman} we consider
$U^{\dagger}(x)i\partial_{\mu}U(x)$. This
object belongs to the Lie algebra of $SU(2)$ and hence can be
decomposed as

\be
U^{\dagger}(x)i\partial_{\mu}U(x) = \amm(x)S_+ + \apm(x)S_- + \bm(x)S^3,
\label{UidU}
\ee
where we have considered the $S_+$ and $S_-$ bases for convenience,
which verifies
\be
   \ba{l}
   S_+ = S^1 + iS^2   \\
   S_- = S^1 - iS^2
   \ea
\quad 
,\quad
   \ba{l}
   [S^3,S_+]=S_+   \cr
   [S^3,S_-]=-S_-  \cr
   [S_+,S_-]=2S^3.
   \ea
\ee

Under the $SU(2)$ transformations
\be
   \ba{ccc}
   \UdidU & \rightarrow  & h(\UdidU)\hd + \partial_{\m}\th S^3
   \ea
\quad , \quad
h = e^{i\theta S^3},
\ee
and hence
\be
   \ba{ccl}
   \amm(x) & \rightarrow & e^{i\th(x)}\amm(x)            \\
   \apm(x) & \rightarrow & e^{-i\th(x)}\apm(x)           \\
   \bm(x)  & \rightarrow & \bm(x) + \partial_{\m}\th(x)..
   \ea
\label{a_SU2_transf}
\ee

Namely, $a^{\pm}_{\mu}$ transforms covariantly whereas $a^3_{\mu}$
transforms like a connexion under an effective $U(1)_{local}$ group,
associated to
the non-linear $SU(2)$ transformations. From now on any reference to the
$U(1)_{local}$ transformation must be understood as the effective
transformation of the non-linear $SU(2)$ transformation over the
fields in (\ref{a_SU2_transf}). These are taken as the basic
building blocks of our construction together with their derivatives. In
order to construct the effective lagrangian a covariant derivative over
$\amm(x)$ can be defined as
\be
D_{\m} \equiv \partial_{\m} - i\bm(x).
\label{a_covariant_derivative}
\ee

Although the connexion does not transform covariantly under $SU(2)$,
an invariant field strength can be constructed in the standard way,
\be
F_{\m\n}(x) \equiv \partial_{\m} a_{\n}^3(x) - \partial_{\n}\bm(x).
\ee

Hence we can in principle construct all invariant terms in the effective 
lagrangian out of $\amm(x), \apm(x), F_{\m\n}(x)$ and $D_{\m}$.
There are however terms which are invariant up to
a total derivative which have to be included in the effective lagrangian. 
These terms are built out of $\bm(x)$ and are usually called topological. 
They read
\be
   \ba{l}
   \bm             \\
   \e^{\m\n\r} \bm \partial_{\n} a_{\r}^3.
   \ea
\label{topological}
\ee

The first term will be important later on.
The second term is the well known abelian Chern-Simons form. As we will
see our space-time symmetries do not allow this term. However it arises
in other condensed matter systems as for instance in quantum Hall
ferromagnets \cite{Ray}.

Given the projector $P_+$ over the higher spin state the
fields in (\ref{UidU}) can be written as
\bea
\amm(x) & = & {1 \over 2s} tr ([\UdidU,S_-]P_+) \nonumber   \\
\apm(x) & = & -{1 \over 2s} tr ([\UdidU,S_+]P_+) \label{a_explicit}\\
\bm(x)  & = &  {1 \over s} tr (\UdidU P_+).   \nonumber
\eea

Once we have an explicit representation for $a_{\m}^{\pm}$ and $\bm$
some relevant properties can be proved, namely,
$F_{\m\n} \sim (\apm a_{\n}^- - a_{\n}^+ \amm)$ and 
$D_{\m} a_{\n}^- = D_{\n} \amm$ (see Appendix B). These properties
ensure that the invariant terms in the effective lagrangian can be 
constructed in terms of $a_{\m}^{\pm}$ and symmetrised covariant 
derivatives acting on them only.

The expert reader may wonder why we do not use the standard $O(3)$
sigma model formulation where the effective lagrangian is built out of
$n^{a}(x)$, an $SU(2)$ vector such that $n^{a}n^{a}=1$. The reason is
simple: any local invariant that
can be constructed in the $O(3)$ sigma formulation can be constructed in
the formulation above (we prove this in appendix~D). However, the
opposite is not true. The essential difference comes from topological terms,
namely, terms that are invariant up to a total derivative. Those are very
elusive in the $O(3)$ sigma model formulation but well under control in
our formulation, as it should be clear from (\ref{topological}).

\subsection{Building blocks: Space-time transformations}
\indent

The most efficient procedure to construct the effective lagrangian is the 
following:

\begin{enumerate}
\item First we construct all the $SU(2)$ invariants under the transformations 
(\ref{a_SU2_transf}), order by order in the derivative expansion.
\item After that we search for invariants under the space-time transformations 
(\ref{U_F_space_transf})-(\ref{U_time_transf}) among those terms.
\end{enumerate}

This procedure allows us to ignore the $SU(2)$ spin transformations induced 
by the space-time transformations in 
(\ref{U_F_space_transf})-(\ref{U_time_transf}). Hence, the effective space
transformations for the ferromagnetic systems read
\be
   \ba{rl}
   \xi:\{ C_{3z}^+,I,\s_y \}: &
      \left\{ \ba{ccc}
      \amm & \rightarrow & a_{\xi\m}^-   \\
      \bm  & \rightarrow & a_{\xi\m}^3,
      \ea \right.
   \ea
\label{a_F_space_transf}
\ee
where the symbol $\xi\m$ stands for the transformed index $\m$ under the
space transformation $\xi$ together with the appropriate coefficient in
each case. Recall that the subindex $\m$ corresponds to a derivative, which 
transforms with the inverse representation of the space points given in 
(\ref{space_transf}). Whereas for the antiferromagnetic systems they become
\beasn
& \phantom{,I}   \xi:\{ C_{3z}^+,\s_y \}:  &
      \left\{ \ba{ccc}
      \amm & \rightarrow & a_{\xi\m}^-  \\
      \bm  & \rightarrow & a_{\xi\m}^3
      \ea \right.                  \\
&               &   \nonumber         \\
& \phantom{C_{3z}^+,\s_y,}   \xi: \{ I \} : &
      \left\{ \ba{ccc}
      \amm & \rightarrow & -a_{\xi\m}^+  \\
      \bm  & \rightarrow & -a_{\xi\m}^3.
      \ea \right.
\label{a_AF_space_transf}
\eeasn

The effective time reversal transformation for both ferromagnet and
antiferromagnet reads
\be
   \ba{rl}
   T: &
      \left\{ \ba{ccc}
      \amm & \rightarrow & -a_{t\m}^+   \\
      \bm & \rightarrow & -a_{t\m}^3,
      \ea \right.
   \ea
\label{a_time_transf}
\ee
again $t\m$ represents the transformation of the index $\m$ under time
reversal symmetry $T$. Let us remark at this point that because
of the different transformation properties under the space symmetries
the effective lagrangians for the ferromagnetic and antiferromagnetic
spin waves will be different. This is in fact not surprising since it is
well known that the low momentum dispersion relation of the spin waves
is quadratic for the ferromagnet but linear for the antiferromagnet. We
shall obtain this result from symmetry considerations only. The
difference arises from the different transformations given by 
(\ref{a_F_space_transf}) for the ferromagnet and by (\ref{a_AF_space_transf})
for the antiferromagnet. In fact only the terms with an odd number of 
time derivatives lead eventually to the above mentioned differences. 
In order to prove this let us consider
the set of generators $\{ C^{+}_{3z} , \sigma_{y} , I , T \}$, as displayed in 
(\ref{a_F_space_transf}) and (\ref{a_time_transf}), for the ferromagnet and 
$\{  C^{+}_{3z} , \sigma_{y} , TI , T \}$ for the antiferromagnet, where we 
choose $TI$ instead of $I$ in (\ref{a_AF_space_transf}b) as a generator. 
Notice that $\{ C^{+}_{3z} , \sigma_{y} , T \}$ act identically in the 
ferromagnetic and antiferromagnetic case, and hence the only differences may 
arise due to action of $I$ or $TI$. Consider next $TI$ on the space 
derivatives $p=z ,\bar z, 3$ for the antiferromagnetic case,
\be
   \ba{rl}
   T\xi:\{ TI \}: &
      \left\{ \ba{ccc}
      a_{p}^- & \rightarrow & a_{\xi p}^-   \\
      a_{p}^3 & \rightarrow & a_{\xi p}^3,
      \ea \right.
   \ea
\ee
and notice that these transformations are identical to those of $I$  on space
derivatives for the ferromagnetic case (\ref{a_F_space_transf}). Finally the 
action of $TI$ on time derivatives for the antiferromagnet reads 
\be
   \ba{rl}
   T\xi:\{ TI \}: &
      \left\{ \ba{ccc}
      a_0^- & \rightarrow & -a_{0}^-   \\
      a_0^3 & \rightarrow & -a_{0}^3.
      \ea \right.
   \ea
\ee
which only differs by a sign from the action of $I$ on time derivatives for
ferromagnets (\ref{a_time_transf}).
Therefore the invariants with an even number of time derivatives are the 
same for both ferromagnetic and antiferromagnetic spin waves. Notice that 
the proof which we carried out is independent of the particular point
group we choose; the only thing we have to do is to substitute the space
transformation $\xi$ which maps points with opposite magnetisation (in
our case the inversion $I$) by itself followed by the time reversal
transformation $T\xi$.

It is worth mentioning at this point that if primitive translations which map 
points with opposite magnetisation existed, terms with an odd number of
time derivatives would not be allowed (see Appendix A). Only in this case
the ground state of the antiferromagnet appears to be time reversal 
invariant at macroscopic scales, which enforces the above restriction on
the effective lagrangian \cite{Burgess}. However, as it should be clear from 
the above, this is not the most general situation, although it is the most
usual one.

\subsection{Derivative expanssion: Power counting}
\indent

The organisation of the effective lagrangian in terms of derivatives is
easier in relativistic theories than in non-relativistic ones. Energy
and momentum are universally related in the former, which allows
to control the derivative expansion by means of a single
dimensionful parameter. This parameter may be thought of as the energy
of the first massive excitation. For non-relativistic theories, energy
and momentum need not fit in any precisely given form, so we may expect
the derivative expansion to be controled by at least two independent
parameters: one with dimensions of energy for the time derivatives and
one with dimensions of momentum for the space derivatives. We may
identify the former as the energy of the first gapped excitation $J$ and the
latter as the typical inverts lattice spacing $1/a$. A natural way to relate
the time derivative expansion to the space derivative expansion arises
once the lowest order terms are written down. It consist of counting the
lowest order in time derivatives as being equally important as the
lowest order in space derivatives, no matter how many derivatives are in
either. This procedure enforces that time and space derivatives
are related in the same fashion as energy and momentum in
the dispersion relation. This is the right way to proceed as far
as there is no external source probing the system (for instance if we
wish to calculate the magnetic susceptibility) or the external source
has a typical energy and momentum compatible with the dispersion
relation. Otherwise the counting should be rearranged according to the
the typical energy and momentum of the external source (this will
be the case when probing the system by electromagnetic radiation). The
latter situation never occurs in relativistic theories because, as
mention before, energy and momentum are universally related.


\section{Effective Lagrangian for the Ferromagnet}
\indent

As we mention in the introduction the effective lagrangian for the Goldstone 
modes must be constructed order by order in space and time derivatives.

The lowest order terms in space derivatives being invariant both under
$SU(2)$ and the space-time symmetries read
\be
   \ba{l}
   \apz \amzb + \amz \apzb \\
   \ap3 \am3.
   \ea
\label{space_a2}
\ee

We call these terms $O(p^2)$. It is very easy to convince oneself that
there are no $O(p^1)$ terms (i.e., with a single space derivative) which
are invariant. However, (\ref{topological}) provides a term which is
invariant up to a total time derivative. It reads
\be
   \ba{l}
   \bo.
   \ea
\label{time_a1}
\ee

This term appears in the literature \cite{Leutwyler,Volovik,Moon} in a
variety of forms, none of which being local as above, which we will show 
to be equivalent to (\ref{time_a1}) in the Appendix C. There, we also 
discuss its relation to certain topological objects.

Equations (\ref{space_a2}) and (\ref{time_a1}) indicate how time derivatives 
must be counted in relation to space derivatives. A time derivative must be 
counted as $O(p^2)$. Let us at this point elaborate a bit on the lowest order 
lagrangian in order to see how it produces the usual dispersion relation for 
ferromagnetic spin waves together with their interaction. We write
\be
{\cal L}(x) = f_{\pi}^2 \left[{1\over 2}\bo -
             {1 \over m}(\apz \amzb + \amz \apzb) -
             {1 \over 2\g m} \ap3 \am3 \right],
\ee
where $f_{\pi}$, $m$ and $\gamma$ are free parameters. The connexion
$\UdidU$ is expanded in terms of the Goldstone modes field,
\be
\UdidU = -{1 \over f_{\pi}^2}\left[
         (f_{\pi} \pa_{\m}\pi^- + \cdots) S_+ +
         (f_{\pi} \pa_{\m}\pi^+ + \cdots) S_- +
     (i(\pi^+\pa_{\m}\pi^- - \pi^-\pa_{\m}\pi^+) + \cdots)S^3\right],
\label{expansion}
\ee
where $\pi^{\pm} = (\pi^1 \pm i\pi^2)/\sqrt{2}$, giving rise to
\be
{\cal L}(x) = \pi^-i\pa_0\pi^+ -{1 \over 2m}\pa_i\pi^-\pa_i\pi^+
                          -{1 \over 2\g m}\pa_3\pi^-\pa_3\pi^+,
\label{Schrodinger}
\ee
up to quadratic order in the spin wave fields. From (\ref{Schrodinger}) 
we see clearly that a time derivative must be counted as $1/2m$ two space 
derivatives. Observe that because (\ref{Schrodinger}) is first order in 
time derivatives, the remaining $U(1)$ spin symmetry implies that the number 
of ferromagnetic spin waves is conserved. It is also remarkable that the 
interaction of any number of spin waves at this order, which is obtained 
by keeping more terms of the expansion (\ref{expansion}) in 
(\ref{Schrodinger}), is given by just three constants, namely $f_{\pi}$, 
$m$ and $\gamma$.

The order of magnitude of the constants above follows from the fact that the
effective lagrangian is an expansion for low energy and momentum controlled by 
the parameters $J$ and $1/a$, namely the energy of the first gapped excitation 
and the typical lattice spacing respectively. $J$ suppresses the time 
derivatives and $1/a$ the space ones. Since the lagrangian density has 
dimensions of $(energy).(momentum)^3$ we can estimate the size of each term 
by writing ${\cal L}(x)\sim J/a^3 \times ({\it dimensionless \, quantities})$ 
and taking into account that the dimensionless quantities are built out of 
time derivatives over $J$ and space derivatives over $1/a$. We obtain 
$f_{\pi}^2\sim 1/a^3$ and $1/m \sim a^2 J$, in accordance with standard 
microscopic calculations in the Heisenberg model.  

The following non-trivial order is $O(p^4)$, which gives rise to the terms 
below:
\bea
& & \ba{l}
   \apo \amo    \\
   i(D_0\ap3 \am3 - D_0\am3 \ap3)    \\
   i[(D_0\apz \amzb - D_0\amz \apzb) -
            (D_0\amzb \apz - D_0\apzb \amz)]  \\
                       \\
    \ea  \label{der_a2}  \\
& & \ba{l}
   \ap3\am3\ap3\am3  \\
   \ap3\am3(\apz\amzb + \amz\apzb)   \\
   \ap3\amz\ap3\amzb + \am3\apzb\am3\apz    \\
   \apz\amz(\ap3\amz + \am3\apz) +
             \amzb\apzb(\am3\apzb + \ap3\amzb)   \\
   \apz\amzb\apz\amzb + \amz\apzb\amz\apzb    \\
   \apz\amzb\apzb\amz  \\
   D_3\ap3 D_3\am3    \\
   D_3\apz D_3\amzb + D_3\amz D_3\apzb     \\
   (D_3\apz D_z\amz + D_3\amz D_z\apz) +
            (D_3\amzb D_{\zb}\apzb + D_3\apzb D_{\zb}\amzb)   \\
   D_z\apzb D_z\amzb + D_{\zb}\amz D_{\zb}\apz.
\ea
\label{space_a4}
\eea
Notice that imposing rotational invariance would reduce (\ref{der_a2}) and 
(\ref{space_a4}) to two and three terms respectively.



\section{Effective Lagrangian for the Antiferromagnet}
\indent

In this section we follow exactly the same logical steps as in the
ferromagnetic case, but taking into account that the transformation
properties of $U(x)$ under the point group are different.

The lowest order terms in space derivatives are exactly the same as in
(\ref{space_a2}). Namely,
\be
   \ba{l}
   \apz \amzb + \amz \apzb  \\
   \ap3 \am3.
   \ea
\label{space_a2_AF}
\ee

The lowest order term in time derivatives is not (\ref{time_a1}) anymore.
Indeed, the transformation properties under $I$ now forbid this term.
Therefore as it was pointed out in section~3 the difference between
ferromagnetic and antiferromagnetic spin waves arises from terms
containing an odd number of time derivatives. Then the lowest order term 
in time derivatives is in this case
\be
   \ba{l}
   \apo \amo.
   \ea
\label{time_a2}
\ee

Hence the effective lagrangian to lowest order reads
\be
{\cal L}(x) = f_{\pi}^2 \left[ \apo \amo -
2v^2(\apz \amzb + \amz \apzb) -  (\g v)^2 \ap3 \am3 \right],
\ee
where $f_{\pi}$ is the spin stiffness, $v$ the spin wave velocity in
$x-y$ plain and $\g v$ the spin wave velocity in the $z$ direction.
In terms of the spin wave fields (\ref{expansion}), the lagrangian above
reads
\be
{\cal L}(x) = \pa_0\pi^+ \pa_0\pi^-
                - v^2 \pa_i\pi^+ \pa_i\pi^-
                - (\g v)^2 \pa_3\pi^+ \pa_3\pi^-,
\label{Klein-Gordon}
\ee
where it is apparent that now we have a linear dispersion relation. It
also becomes apparent that the time derivatives must be counted as $v$
times a space derivative. Observe that because (\ref{Klein-Gordon}) is
second order in derivatives it describes two degrees of freedom. The 
remaining $U(1)$ spin symmetry tells us that antiferromagnetic spin waves 
can only be produced (or annihilated) in pairs. This is completely analogous 
to a relativistic theory: one of the spin waves plays the role of a (massless) 
particle and the other of an antiparticle. The total number of particles plus 
antiparticles is conserved  due to the $U(1)$ symmetry. It is again
remarkable that the interaction between spin waves at this order, which 
we would obtain by keeping further terms in the expansion (\ref{expansion}), 
is given in terms of three parameters only, namely $f_{\pi}$, $v$ and $\gamma$
\cite{Hasenfratz}.

The order of magnitude of the constants above is estimated as in the 
ferromagnetic case. Now we obtain  $f_{\pi}^2\sim 1/Ja^3$ and $v\sim Ja$, 
which is again in accordance with microscopic calculations in the Heisenberg 
model.

It is worth mentioning that the following extra term appears at the 
lowest order
\be
F_{03} \sim (a_{0}^+ a_{3}^- - a_{0}^-a_{3}^+ ).
\ee
This term is a total derivative and will be dropped. This is fine as far as 
we stay within a perturbative approach. However, since it is a total 
derivative of an object which is not $SU(2)$ invariant, it may become 
relevant if a non-perturbative analysis is attempted. 

The next to leading non-trivial order is $O(p^4)$, which reads
\bea
& & \ba{l}
   \apo\amo\apo\amo    \\
   i\apo\amo(\apo\am3 - \amo\ap3)   \\
   \apo\amo\ap3\am3    \\
   \apo\am3\apo\am3 + \amo\ap3\amo\ap3   \\
   i(\apo\am3 - \amo\ap3)\ap3\am3   \\
   \apo\amo(\apz\amzb + \amz\apzb)  \\
   \apo\amz\apo\amzb + \amo\apzb\amo\apz  \\
   i(\apo\am3 - \amo\ap3)(\apz\amzb + \amz\apzb)   \\
   i(\apo\amz\ap3\amzb - \amo\apzb\am3\apz)   \\
   i[\apz\amz(\apo\amz - \amo\apz) -
             \amzb\apzb(\amo\apzb - \apo\amzb)]  \\
   D_0\apo D_0\amo  \\
   D_0\ap3 D_0\am3  \\
   D_0\apz D_0\amzb + D_0\amz D_0\apzb \\
   i[D_0\ap3(D_z\amzb + D_{\zb}\amz) -
             D_0\am3 (D_{\zb}\apz + D_z\apzb)]  \\
   \ea
\label{space_AF_a4}
\eea
together with the space derivatives terms given in (\ref{space_a4}). Notice
that the above terms with an odd number of time derivatives would not appear
if a primitive translation mapping points with opposite magnetisations 
existed in the $Cr_2O_3$ (see Appendix A).
Notice also that imposing rotational invariance would reduce 
(\ref{space_AF_a4}) to five terms.


\section{Spin-orbit Corrections}
\indent

The spin-orbit coupling of the electrons in the magnetic ions is usually
the main source of explicit $SU(2)$ breaking. This explicit breaking can
be taken into account in Heisenberg-type models by the inclusion of
two new terms in the hamiltonian \cite{Moriya}
\be
H = \sum_{<i,j>} J_{ij}\Sv_i \Sv_j +
    \sum_{<i,j>} \Dv_{ij}(\Sv_i \times \Sv_j) +
    \sum_{<i,j>} M_{ij}^{ab} S^{a}_i S^b_j,
\label{Moriya}
\ee
where $D^a \sim (\D g /g)J$ and $M^{ab} \sim (\D g /g)^2J$ are the 
antisymmetric and symmetric any\-so\-tro\-pies respectively ($M^{ab}$ is 
symmetric with respect the spin indices). These are related to the
(super)exchange coupling through the change in the effective gyromagnetic
factor due to the spin-orbit interaction.. Typically $\D g \sim 10^{-2}g$ 
\cite{libros}.

In order to incorporate the effects of (\ref{Moriya}) in the effective
theory we promote $\Dv_{ij}$ and $M_{ij}^{ab}$ to source fields 
$D^a(\xv_i,\xv_j)$ and $M^{ab}(\xv_i,\xv_j)$ and assign transformation 
properties to them such that (\ref{Moriya}) becomes $SU(2)$ invariant,
\be
   \ba{rcl}
   D^a(\xv_i,\xv_j) & \rightarrow & R^a_{\ b} D^b(\xv_i,\xv_j)  \\
   M^{ab}(\xv_i,\xv_j) & \rightarrow & R^a_{\ c}R^b_{\ d}
   M^{cd}(\xv_i,\xv_j).
   \ea
\ee

Now, if we could derive our effective lagrangian for the spin waves from the 
microscopic model it would be a functional of $U(x)$, $D^a$ and $M^{ab}$
invariant under $SU(2)$ transformations. Once we particularise the sources to
reproduce the anisotropic spin-orbit tensors we automatically obtain
the effects of the latter in the effective theory.

Since the spin-orbit corrections lead to short range interactions in
(\ref{Moriya}), the local limit of the sources will be taken.
In this limit the leading contribution compatible with the crystal 
symmetries of the antisymmetric anisotropy is represented by a tensor 
with one $SU(2)$ and two symmetric space indices, $D_{pq }^a$, where 
$p,q=z,\bar z ,3$ (these indices transform with the inverse representation of
the space points (\ref{space_transf})). The symmetric tensor is represented
by a second order tensor of $SU(2)$ with no space indices, $M^{ab}$.

Let us next consider the following objects which transform covariantly
under $SU(2)$,
\be
   \ba{rcl}
   D_{pq } \equiv D_{pq }^aS^a  & \rightarrow  & gD_{pq }g^{\da}  \\
   M \equiv M^{ab}(S^a \otimes S^b + S^b \otimes S^a)   & \rightarrow  &
          (g \otimes g)M(g^{\da} \otimes g^{\da}).
   \ea
\ee

When these sources are set to their most general form compatible with
the crystal symmetries only a few non-vanishing terms remain. Namely,

\bea
   &  \ba{l}
      D_{zz} = D_{zz}^- S_+         \\
      D_{\zb\zb} = D_{\zb\zb}^+ S_-
      \ea
   \qquad \quad D_{zz}^- = -D_{\zb\zb}^+ &  \nn  \\
   &   &  \nn    \\
   &  \ba{l}
      D_{3z} = D_{3z}^+ S_-        \\
      D_{3\zb} = D_{3\zb}^- S_+
      \ea
   \qquad \quad D_{3z}^+ = -D_{3\zb}^-   &  \label{DM}   \\
   &   &  \nn    \\
   &  M = M^{-+}(S_+ \otimes S_- + S_- \otimes S_+) + 
          M^{33}(S^3 \otimes S^3), &  \nn
\eea
where $ D_{\zb\zb}^+$, $ D_{3z}^+$,  $ M^{-+}$ and $ M^{33}$ are free
parameters.

\subsection{Spin-orbit sources: Internal transformations}
\indent

Since our basic building blocks transform in a simple way under
the $U(1)_{local}$ in the $SU(2)$ non-linear realisation, it is convenient
to introduce new sources with simple transformation properties under
$U(1)_{local}$  in the following way:
\be
U^{\da}(x) D_{pq } U(x) = \dmmn(x)S_+ + \dpmn(x)S_- + \emn(x)S^3.
\label{UDU}
\ee

These new sources transform as
\be
   \ba{ccl}
   \dmmn(x) & \rightarrow  & e^{i\th(x)}\dmmn(x)   \\
   \dpmn(x) & \rightarrow  & e^{-i\th(x)}\dpmn(x)  \\
   \emn(x)  & \rightarrow  & \emn(x).
   \ea
\ee

The same can be done for the symmetric source,
\bea
\lefteqn{ \left(U^{\da}(x) \otimes U^{\da}(x)\right) M
               \Bigl( U(x) \otimes U(x) \Bigr) =
                           \mmm(x)(S_+ \otimes S_+) } \nonumber  \\
 & & \phantom{ (U^{\da}(x) \otimes U^{\da}(x))MU}
     \mbox{} + \mpp(x)(S_- \otimes S_-)  \nonumber \\
 & & \phantom{ (U^{\da}(x) \otimes U^{\da}(x))MU}
     \mbox{} + \md3(x)(S^3 \otimes S^3)  \nonumber \\
 & & \phantom{ (U^{\da}(x) \otimes U^{\da}(x))MU}
     \mbox{} + \mmp(x)(S_+ \otimes S_- + S_- \otimes S_+)
                                               \label{UUMUU} \\
 & & \phantom{ (U^{\da}(x) \otimes U^{\da}(x))MU}
     \mbox{} + \mm3(x)(S_+ \otimes S^3 + S^3 \otimes S_+) \nonumber \\
 & & \phantom{ (U^{\da}(x) \otimes U^{\da}(x))MU}
     \mbox{} + \mp3(x)(S_- \otimes S^3 + S^3 \otimes S_-). \nonumber
\eea

The transformation properties for these components are
\be
   \ba{lcl}
   \mmm(x) & \rightarrow & e^{2i\th(x)} \mmm(x) \\
   \mpp(x) & \rightarrow & e^{-2i\th(x)} \mpp(x) \\
   \mm3(x) & \rightarrow & e^{i\th(x)} \mm3(x) \\
   \mp3(x) & \rightarrow & e^{-i\th(x)} \mp3(x) \\
   \mmp(x) & \rightarrow & \mmp(x)    \\
   \md3(x) & \rightarrow & \md3(x).
   \ea
\ee

An explicit representation for the sources $d_{pq }^a(x)$ introduced in
(\ref{UDU}) is given by
\bea
\dmmn(x) & = &  {1 \over 2s} tr ([U^{\da}D_{pq }U,S_-]P_+) \nonumber \\
\dpmn(x) & = & -{1 \over 2s} tr ([U^{\da}D_{pq }U,S_+]P_+)
                                               \label{d_explicit} \\
\emn(x)  & = &  {1 \over s} tr (U^{\da}D_{pq }U P_+),      \nonumber
\eea
and a similar but lengthier expression can be given for $m^{ab}(x)$ in
(\ref{UUMUU}). From the explicit representation (\ref{d_explicit}) for
the sources $d_{pq }^a$ together with (\ref{DM}) it is obvious that
\mbox{$d_{zz}^a d_{3\zb}^b = d_{zz}^b d_{3\zb}^a$},
\mbox{$d_{\zb\zb}^a d_{3z}^b = d_{\zb\zb}^b d_{3z}^a$} and
\mbox{$d_{zz}^a d_{3z}^b = d_{\zb\zb}^b d_{3\zb}^a$}.
Moreover one can easily prove that derivatives over these sources can be 
written as the source itself multiplied by $a_{\m}^{\pm}(x)$ or
$a_{\m}^3(x)$. Therefore only $d_{pq }^a(x)$ and $m^{ab}(x)$ and not
their derivatives have to be used in addition to our basic building blocks 
(\ref{UidU}) to construct the effective lagrangian.

\subsection{Spin-orbit sources: Space-time transformatinos}
\indent

Let us next see how these sources transform under the space-time symmetries. 
Recall first that a space transformation induces a $SU(2)$ spin transformation 
also in the sources. However, as it was pointed out in section 3, the $SU(2)$
transformations induced by space-time transformations can be ignored because 
the lagrangian is first constructed to be $SU(2)$ invariant.


For the ferromagnet the effective $\bar3m$ transformations are given by
\be
   \ba{rl}
   \xi:\{ C_{3z}^+,I, \s_y\}: &
      \left\{ \ba{ccc}
      \dmmn  & \rightarrow & d_{\xi p \xi q }^-    \\
      \emn   & \rightarrow & d_{\xi p \xi q }^3    \\
      m^{ab} & \rightarrow & m^{ab},
      \ea \right.
   \ea
\label{d_F_space_transf}
\ee
and for the antiferromagnet
\beasn
& \phantom{,I}   \xi:\{ C_{3z}^+,\s_y\}: &
      \left\{ \ba{ccc}
      \dmmn  & \rightarrow & d_{\xi p \xi q }^-    \\
      \emn   & \rightarrow & d_{\xi p \xi q}^3    \\
      m^{ab} & \rightarrow & m^{ab}
      \ea \right.
                             \\
&             &   \nonumber          \\
& \phantom{C_{3z}^+,\s_y,}   \xi:\{ I \}: &
      \left\{ \ba{ccc}
      \dmmn & \rightarrow & -d_{\xi p \xi q }^+     \\
      \emn  & \rightarrow & -d_{\xi p \xi q }^3    \\
      \mmm  & \rightarrow & \mpp                  \\
      \mmp  & \rightarrow & \mmp                  \\
      \mm3  & \rightarrow & \mp3                 \\
      \md3  & \rightarrow & \md3.
      \ea \right.
\label{d_AF_space_transf}
\eeasn

Time reversal, like in (\ref{a_time_transf}), gives the same
transformations both for the ferromagnet and the antiferromagnet,
\be
   \ba{rl}
   T: &
      \left\{ \ba{ccc}
      \dmmn & \rightarrow & - \dpmn    \\
      \emn  & \rightarrow & - \emn     \\
      \mmm  & \rightarrow & \mpp       \\
      \mmp  & \rightarrow & \mmp       \\
      \mm3  & \rightarrow & \mp3       \\
      \md3  & \rightarrow & \md3.
      \ea \right.
   \ea
\label{d_time_transf}
\ee

Again, as in section 3, it is easy to prove that the transformations
(\ref{d_F_space_transf}) and (\ref{d_time_transf}) give the same
invariants as (\ref{d_AF_space_transf}) and
(\ref{d_time_transf}). Consider instead of (\ref{d_AF_space_transf}b)
the transformation given by $T\xi$,
\be
   \ba{rl}
   T\xi:\{ TI \}: &
      \left\{ \ba{ccc}
      \dmmn  & \rightarrow & d_{\xi p \xi q }^-    \\
      \emn   & \rightarrow & d_{\xi p \xi q }^3    \\
      m^{ab} & \rightarrow & m^{ab}.
      \ea \right.
   \ea
\ee

Therefore all the invariants constructed out of $d$s, $m$s and $a$s with an 
arbitrary number of space derivatives and an even number of time derivatives 
are the same for ferromagnetic and antiferromagnetic spin waves.

\subsection{Invariant terms}
\indent

The previous transformation properties lead to the following
invariants at leading order in terms of $d$s and $m$s,

\be
   \ba{l}
   \dpzz \dmzbzb + \dmzz \dpzbzb \\
   (\dpz3 \dmzz + \dmz3 \dpzz) + (\dmzb3 \dpzbzb + \dpzb3 \dmzbzb) \\
   \dpz3 \dmzb3 + \dmz3 \dpzb3  \\
   \ezz \ezbzb  \\
   \ez3 \ezz + \ezb3 \ezbzb \\
   \ez3 \ezb3  \\
               \\
   \mmp   \\
   \md3.
   \ea
\label{d2}
\ee

If we expand the
$d$s and $m$s sources in terms of the spin wave
fields we obtain
\bea
U^{\da}D_{pq}U & = & \left[ D_{pq}^-
      + {1 \over f_{\pi}^2}(D_{pq}^+\pi^-\pi^+ - D_{pq}^-\pi^+\pi^-)
      + \cdots \right] S_+   \nonumber \\
 & + &    \left[ D_{pq}^+
      + {1 \over f_{\pi}^2}(D_{pq}^-\pi^+\pi^+ - D_{pq}^+\pi^+\pi^-)
      + \cdots \right] S_-    \\
 & + &    \left[ -{2i \over f_{\pi}}(D_{pq}^+\pi^- -
                        D_{pq}^-\pi^+) + \cdots \right]S^3, \nn
\eea
\bea
\lefteqn{ (U^{\da} \otimes U^{\da}) M (U \otimes U) =
     [(2M^{-+} -M^{33})\pi^-\pi^- + \cdots ]
                 (S_+ \otimes S_+)}  \qquad \qquad \nonumber \\
 & & \mbox{} + [(2M^{-+} -M^{33})\pi^+\pi^+ + \cdots ]
                 (S_- \otimes S_-) \nonumber \\
 & & \mbox{} + [M^{-+} - 2(2M^{-+} -M^{33})\pi^+\pi^- + \cdots ]
                 (S_+ \otimes S_- + S_- \otimes S_+) \nonumber \\
 & & \mbox{} + [-i(2M^{-+} -M^{33})\pi^- + \cdots ]
                 (S_+ \otimes S^3 + S^3 \otimes S_+) \\
 & & \mbox{} + [i(2M^{-+} -M^{33})\pi^+ + \cdots ]
                 (S_- \otimes S^3 + S^3 \otimes S_-) \nonumber \\
 & & \mbox{} + [2(2M^{-+} -M^{33})\pi^+\pi^- + M^{33} + \cdots ]
                 (S^3 \otimes S^3).  \nonumber
\eea

 From these two expansions we can easily see that the terms (\ref{d2})
at quadratic order in the spin waves fields produce a gap in the dispersion 
relation. Notice however that the energy gap for the spin waves is $\sim d^2$ 
for the ferromagnet, since its spin waves verify a Schr\"odinger like 
equation, whereas it is $\sim d$ for the antiferromagnet, since its spin 
waves verify a Klein-Gordon type equation.

Notice also that the terms in (\ref{d2}) contribute to the ground state energy 
and force the (staggered) magnetisation to be along the third direction. This 
is why we took the direction of the spontaneously symmetry breaking along the 
third axes in (\ref{spin_waves}). Otherwise terms with a single spin wave 
field would appear in the expansion of (\ref{d2}) indicating us that we chose 
a wrong direction for the (staggered) magnetisation.


The above considerations allow us to decide how to take the relative
order of magnitude for the derivatives and the anisotropy tensors.
If we consider, for instance, the typical energy of the spin waves of
the order of their energy gap, the
counting for the ferromagnet becomes $a_0 \sim a_i^2 \sim d^2 \sim m$,
which leads to the following $O(p^3)$ terms:
\be
   \ba{l}
   i[\ezbzb(\apzb\am3 - \amzb\ap3) - \ezz(\amz\ap3 - \apz\am3)] \\
   i[\ez3(\apzb\am3 - \amzb\ap3) - \ezb3(\amz\ap3 - \apz\am3)],
   \ea
\label{ad3}
\ee
whereas for the antiferromagnet an analogous counting implies
$a_0^2 \sim a_i^2 \sim d^2 \sim m$, which in addition to the previous 
(\ref{ad3}) terms leads to new $O(p^3)$ terms:
\be
   \ba{l}
   \ezbzb(\apzb\amo + \amzb\apo) + \ezz(\amz\apo + \apz\amo) \\
   \ez3(\apzb\amo + \amzb\apo) + \ezb3(\amz\apo + \apz\amo).
   \ea
\ee
Notice again that if primitive translations which map points with
opposite magnetisations existed these terms would be forbidden.



\section{Magnetic Dipole Corrections}
\indent

A second source of explicit $SU(2)$ breaking are the magnetic dipole
interactions. They have the form \cite{libros}
\be
H = \m^2 \sum_{i \neq j} \left(
{\Sv_i\Sv_j \over |\xv_i - \xv_j|^3} -
    {(\Sv_i\hat\xv_i) (\Sv_j\hat\xv_j) \over |\xv_i - \xv_j|^3}
\right).
\ee

The first term is $SU(2)$ invariant, and has the form of the
(super)exchange parameter in the Heisenberg model,
while the second term has the form of the
symmetric anisotropy in (\ref{Moriya}),
\be
{1 \over |\xv_i - \xv_j|^3} \sim J_{ij}
\qquad , \qquad
{\hat x_i^a \hat x_j^b \over |\xv_i - \xv_j|^3} \sim M^{ab}_{ij},
\ee

Let us point out however that whereas the superexchange and anisotropic
terms lead to short range interactions the terms above lead to long range
interactions. In spite of this, the local limit will be taken since the
strength of the interactions
decays like the third power of the distance between the magnetic dipoles
and furthermore it is usually very small compared to the spin-orbit terms. 
In fact this long range interaction starts playing a crucial role at very 
long distance phenomena, like in the formation of domain walls in
ferromagnets \cite{libros}, which lie beyond the scope of this work.
If we restrict ourselves to the local limit, then both terms
have been already taken into account in the $SU(2)$ invariant structure
$(J_{ij})$ and in the explicit symmetry breaking source $(M^{ab})$.



\section{The Coupling to Electromagnetic Fields}
\indent

If we probe our ferromagnet or antiferromagnet by electromagnetic waves
the energy of which is much smaller than the energy of the first
gaped excitation, the only relevant magnetic degrees of freedom are the
spin waves. It is then relevant how to include the electromagnetic
fields in the effective lagrangians built in the previous sections.

When the electromagnetic fields enter the game the $U(1)_{em}$  local gauge
invariance is the additional symmetry that we have to take into account.
Since the fields $U(x)$ have trivial transformation properties under the
$U(1)_{em}$ (spin waves have no electric charge) we may na\"{\i}vely expect
non-minimal couplings (couplings to the field strength tensor) only. However,
the Pauli term which arises in any microscopic model when a magnetic
field is present  breaks explicitly the $SU(2)$ spin symmetry in a very
particular way. We shall first address how to introduce in the effective
lagrangian the effects of the microscopic Pauli term.

\subsection{Pauli coupling}
\indent

For definiteness we may have in mind the Heisenberg model, but the argument 
we present goes through for more complicated models like the t-J model, Hubbard
model, etc.. For any microscopic model we can think of the hamiltonian in the 
presence of a magnetic field will be augmented, at least, by the Pauli term, 
which explicitly breaks the $SU(2)$ spin symmetry,
\be
H = \sum_{<i,j>} J_{ij} \Sv_i \Sv_j - \m \sum_i \Sv_i {\bf B}.
\label{Pauli}
\ee

Let us assume that the remaining part of the hamiltonian is invariant under 
the original $SU(2)$ symmetry, as in (\ref{Pauli}), even when the 
electromagnetic fields are switched on. Then the only term which breaks 
this symmetry is the Pauli term. Let us then write the lagrangian in the 
second quantisation formalism,
\be
L =\sum_i \psi^{\da}(x_i)i\pa_0\psi(x_i) +
   \m \sum_i \left(\psi^{\da}(x_i) \Sv \psi(x_i)\right) {\bf B}(x_i)
                                                        + \cdots.
\ee

This lagrangian can be rewritten in a form that the Pauli term is
associated to the time derivative,
\be
L = \sum_i \psi^{\da}(x_i)
        i \left(\pa_0 -i\m\Sv{\bf B}(x_i)\right) \psi(x_i) + \cdots.
\ee

Written in this way the term inside the parenthesis has the form of a
covariant derivative for time dependent $SU(2)$ transformations. Then
if we promote the Pauli term to a new source, $A_0(x) \sim \m \Sv{\bf B}(x)$  
such that it transforms like a connexion under time dependent $SU(2)$
transformations,
\be
A_0(x) \rightarrow g(t)A_0(x)g^{\dagger}(t)
                    + ig(t)\partial_0g^{\dagger}(t),
\ee
the microscopic lagrangian becomes invariant under time dependent $SU(2)$ spin
transformations. Now we can construct the effective theory with this source, 
and finally set it to its actual value. This is completely analogous to the 
procedure carried out in section~6 for the spin-orbit breaking terms. The 
effective theory derived from the microscopic level would be a functional 
of the $U(x)$ and the connexion $A_0(x)$ invariant under time dependent 
$SU(2)$ transformations. This is easily achieved by replacing the time 
derivatives acting on $U$ by covariant time derivatives. Namely,
\be
\partial_0 \rightarrow D_0 \equiv \partial_0 -iA_0(x).
\label{covariant_derivative}
\ee

When we particularise the $A_0(x)$ to reproduce the Pauli term we
obtain its effects in the effective theory. From now on we will have
to write $a_0^{\pm}$ and $a_0^3$ as
\bea
\lefteqn{ U^{\da}iD_0 U = -{1 \over f_{\pi}^2} \left\{
  \left[ f_{\pi}\pa_0\pi^- -
   \m \left( {1 \over 2} (f_{\pi}^2 - \pi^+\pi^-)B^{\zb} +
          {1 \over 2} \pi^-\pi^- B^z + i f_{\pi} \pi^- B^3 \right) +
           \cdots \right] S_+ \right. } \nonumber \\
 & & \mbox{} + \left[ f_{\pi}\pa_0\pi^+ -
      \m \left( {1 \over 2} \pi^+\pi^+ B^{\zb} +
          {1 \over 2} (f_{\pi}^2 - \pi^+\pi^-)B^z -
          i f_{\pi} \pi^+ B^3 \right) + \cdots \right] S_-
                                          \label{Pauli_expansion} \\
 & & \mbox{} + \left. \biggl[ i (\pi^+\pa_0\pi^- - \pi^-\pa_0\pi^+) -
      \m \biggl( i f_{\pi} \pi^+ B^{\zb} -i f_{\pi}\pi^- B^z +
      (f_{\pi}^2 - 2\pi^+\pi^-) B^3 \biggr) + \cdots \biggr] S^3
                                              \right\}, \nonumber
\eea
whereas $a_{i}^{\pm}$ and $a_{i}^3$ are still given by (\ref{expansion}).

It is important to realise that we do not have to introduce any additional
unknown constant, but the effective gyromagnetic factor of the spin
degrees of freedom in the microscopic theory in the above terms. Notice,
however, that space derivatives on $A_0(x)$ transform covariantly under
time dependent $SU(2)$ transformations. Hence they may be used to
construct further invariants in analogy to the $d$s in (\ref{UDU}) and
(\ref{d_explicit}). We will not present these invariants explicitly since, 
as we will see later on, they are very suppressed in realistic situations.

\subsection{Non-minimal couplings}
\indent

Let us next discuss the non-minimal couplings. This is the only way the 
electromagnetic field can couple to non charged excitations like the spin 
waves, namely, by means of the field strength tensor, i.e., electric field, 
${\bf E}$, and magnetic field, ${\bf B}$. The transformation properties 
associated to the electromagnetic fields are the following:
\begin{enumerate}
\item they both are scalars under $SU(2)$ spin transformations,
\item the electric field transforms like a vector under $\bar3m$,
\item the magnetic field transforms like a pseudovector under $\bar3m$, and
\item they both are scalars under under primitive translations,
\item under the time reversal they transform as
\end{enumerate}
\be
   \ba{rl}
   T: &
      \left\{ \ba{ccc}
      E^a  & \rightarrow  &  E^a  \\
      B^a  & \rightarrow  &  -B^a.
      \ea \right.
   \ea
\ee

\subsection{Power counting}
\indent

Once we have the transformation properties for the electromagnetic
fields let us consider their relative size in order to write the
derivative expansion in the effective theory.

The expression (\ref{covariant_derivative}) indicates us that the Pauli term, 
$\m {\bf B}$, has to be suppressed by the first gapped excitation, $J$. 
Consider next the electromagnetic fields in terms which are $SU(2)$ invariant.
They may arise from a microscopic model in two ways: (i) originated by
minimal couplings of the electromagnetic potentials or (ii) by explicit 
non-minimal couplings in the microscopic model, which may arise when 
integrating out even higher scales of energy and momentum. The second kind 
of terms are suppressed by a higher energy and momentum scale and will be 
ignored as far as estimates are concerned. They would slightly modify the 
values of the parameters in the effective lagrangian, which are anyway 
unknown. The relative suppression of the terms originated by minimal coupling 
are simple to estimate: the scalar potential goes always accompanying a time 
derivative and hence it will be suppressed by~$J$, whereas the vector 
potential is associated to a link variable (in a lattice model) and hence 
suppressed by $1/a$ (in the continuum limit the space derivative is associated 
to the link and therefore the vector potential is associated to the space 
derivative). Consequently the electric fields are suppressed by $J/ea$ whereas 
the magnetic fields by $1/ea^2$, where $e$ is the electron charge.

When the electromagnetic field enters the system it fixes the relevant scales 
of energy and momentum. For the effective theory to make sense these must be
much smaller than $J$ and $1/a$ respectively. The amplitude of the 
electromagnetic field is also constrained so that we can organise our 
effective lagrangian in increasing powers of $E$ and $B$ as well as 
derivatives. This requires $\m B / J$, $eaE / J$ and $ea^2B$ be much smaller 
than $1$.

Nothing else can be said in general about the organisation of the effective 
lagrangian. Let us at this point introduce $v=Ja$ which combines the energy 
and momentum scales in a single parameter. This parameter allows us to write 
our theory in terms of a single scale, namely, $J$. For the antiferromagnetic 
spin waves $v$ has a precise meaning: it is the velocity of propagation of 
the spin waves, whereas for the ferromagnetic spin waves it does not have 
any particular meaning. Given typical values for $J \sim 10 meV$ and 
$a \sim 10 \AA$ \cite{libros}, then $v \sim 10^{-4}$ (Recall that we have 
taken $c=1$, therefore $v$ has to be thought as $v/c$).

In order to further illuminate the construction of the effective lagrangian, 
let us consider the interaction with a classical monochromatic electromagnetic 
wave of energy $\omega$. In this case the amplitude of the electric and 
magnetic fields is the same. Let us choose $\omega\sim 10^{-1} J$, which 
tells us that we must count $\partial_0$ as $10^{-1}J$. The momentum of the 
electromagnetic field is in this case $\omega$, which tells us that for 
typical lattice spacings we must count $v \partial_i \sim 10^{-4}\partial_0$. 
Notice that the counting is very different in the absence of electromagnetic 
field. We still have to fix the amplitude of the electromagnetic field. For
simplicity let us choose $eaE\sim 10^{-1}J$. Then the following relative 
suppressions hold:
\be
   \ba{ccc}
   \pa_0, eaE     & \sim  & 10^{-1}J   \\
   d              & \sim  & 10^{-2}J   \\
   \m B, m        & \sim  & 10^{-4}J   \\
   v\pa_i, eavB   & \sim  & 10^{-5}J.
   \ea
\label{suppressions}
\ee
where a typical value for the gyromagnetic factor 
$\m \sim \m_B = 9.27\,10^{-24} J/T$ has been taken. With the counting 
above the following terms are obtained for the ferromagnetic systems:

First order $O(p^1)$:
\be
   \ba{l}
   \bo.
   \ea
\label{FO(p1)}
\ee

Second order $O(p^2)$:
\be
   \ba{l}
   \apo\amo      \\
   E^zE^{\zb}  \\
   E^3E^3.
   \ea
\label{FO(p2)}
\ee

Fourth order $O(p^4)$:
\be
   \ba{l}
   \apo\amo\apo\amo   \\
                  \\
   E^zE^{\zb}E^zE^{\zb}   \\
   E^zE^{\zb}E^3E^3  \\
   E^3E^3E^3E^3  \\
                 \\
   \pa_0E^z\pa_0E^{\zb}  \\
   \pa_0E^3\pa_0E^3  \\
                 \\
   \apo\amo E^zE^{\zb}  \\
   \apo\amo E^3E^3  \\
               \\

   \dpzz \dmzbzb + \dmzz \dpzbzb \\
   (\dpz3 \dmzz + \dmz3 \dpzz) + (\dmzb3 \dpzbzb + \dpzb3 \dmzbzb) \\
   \dpz3 \dmzb3 + \dmz3 \dpzb3  \\
   \ezz \ezbzb  \\
   \ez3 \ezz + \ezb3 \ezbzb \\
   \ez3 \ezb3  \\
               \\
   \mmp   \\
   \md3.
   \ea
\label{FO(p4)}
\ee

Notice that it is not
up to fourth order that we obtain a coupling of the electric field
to the spin waves. The Pauli coupling is encoded in the $a_0$
blocks as given in (\ref{Pauli_expansion}). At $O(p^4)$
there is only a contribution arising from (\ref{FO(p1)}) when we
substitute the time derivative by the covariant derivative
(\ref{covariant_derivative}).

For the antiferromagnet there is no invariant at $O(p^1)$. Second order 
($O(p^2)$) invariants are the same as those for the ferromagnet in 
(\ref{FO(p2)}). At fourth order ($O(p^4)$) in addition to the terms given 
in (\ref{FO(p4)}) we have two new terms,
\be
   \ba{l}
   i[(\dpzbzb\amo - \dmzbzb\apo)E^z - (\dmzz\apo -\dpzz\amo)E^{\zb}] \\
   i[(\dpz3\amo - \dmz3\apo)E^z - (\dmzb3\apo - \dpzb3\amo)E^{\zb}].
   \ea
\ee

Notice that these new terms which appear in the antiferromagnetic
case at $O(p^4)$ contain time derivatives and the spin-orbit tensor which 
breaks explicitly the $SU(2)$ spin symmetry, which give rise to an 
electromagnetic coupling mediated by the spin-orbit interaction. These 
terms would be forbidden if a primitive translation mapping points with 
opposite magnetisations existed.


\section{Summary and Discussion}
\indent

We have provided a systematic way to write down effective lagrangians for spin 
waves in ferromagnetic and antiferromagmetic crystalline solids. We have done 
so at next to leading order. This is achieved by fully exploiting the internal 
symmetry breaking pattern $SU(2)\rightarrow U(1)$ as well as the space 
symmetries and time reversal. We have shown how to introduce explicit
symmetry breaking terms, as those induced by atomic spin-orbit couplings 
and magnetic dipole interactions. We have also shown how to introduce 
couplings to electromagnetic fields.

The alert reader may wonder about the role that the magnetic group plays
in our construction. In fact, it does not play any role else that indicating 
us the symmetry group of the ground state configuration. Notice that when we
have spontaneous symmetry breaking the magnetic group should not be
important since in a first approximation the local magnetisation direction 
is arbitrary, and each direction corresponds to a different magnetic group. 
It is only after introducing explicit $SU(2)$ breaking corrections, like 
those induced by the spin-orbit interactions, that the local magnetisations 
take the direction of a crystal symmetry axis, giving rise to a non-trivial 
magnetic group. If we introduce a constant magnetic field in an arbitrary 
direction whose contribution is larger than that of the spin-orbit
interactions the local magnetisations will point to the direction of the 
magnetic field. Then the magnetic group will be trivial again. Within our 
formalism all the possible situations are taken into account once we have 
assign the proper space-time transformations to the field $U(x)$ in 
(\ref{space_transf})-(\ref{U_AF_space_transf}).

A word of caution is needed when using the effective lagrangians to higher 
orders: quantum (loop) corrections to lower order terms give in general 
contributions of some higher order. Then in order to make a consistent 
calculation to a given higher order, in general a loop calculation at lower 
order is necessary. For the antiferromagnet the kind of loop calculations to be
carried out is completely analogous to calculations in relativistic theories, 
for which there is abundant experience \cite{Gasser,Hasenfratz}. Typically a 
one loop calculation at $O(p^2)$ gives $O(p^4)$, a two loop one $O(p^6)$, etc.
For the ferromagnet, there are no explicit loop calculations in the literature 
to our knowledge. On general grounds, we expect a similar pattern though the 
exact way in which lower order loop contributions combine into higher orders 
might be different{\footnote{A recent non-trivial loop calculation can be 
found in \cite{Hofmann}}}.

We would like to point out a few issues in our work that we find particularly 
relevant.
\begin{enumerate}
\item Our formulation easily keeps track of elusive topological terms.
\item We find the remarkable differences between the ferromagnetic and
antiferromagnetic spin wave effective lagrangian by the only use of symmetry 
properties.
\item Within the antiferromagnetic case, we point out that the existence  
of primitive translations which map points with opposite magnetisations has
non-trivial consequences in the effective lagrangian. 
\item We easily see that the Pauli term does not introduce any new
parameter in the effective lagrangian.
\end{enumerate}

We would like to stress again that an important advantadge of the effective 
lagrangian is that it is model independent. Any microscopic model with the 
same space-time group undergoing the same symmetry breaking pattern gives 
rise to the same effective lagrangian. Only the particular values of the 
constants may differ. However, in view of the relative large number of such 
unkwown constants that arise at next to leading order in our lagrangians, 
one may wonder if they can actually be of any use at all.
We are confident that they will. In particular, 
we propose to use them as a guideline on possible interesting phenomena due to 
spin waves. If one suspects that a given material may have exciting magnetic 
properties in the spin wave region, one need not carry out a microscopic 
calculation to check so. By writing down the effective lagrangian one can 
most easily check if the expected phenomenon may occur or not. To do this 
one has to input realistic numbers for the $J$ and $a$ parameters according 
to the given material, changing the counting (\ref{suppressions}) if
necessary, and proceed as above. In case it is feasible, a microscopic 
calculation may be supplemented to fix the unknown parameters. Recently, 
we have presented a non-trivial application of these techniques  
in \cite{Soto}. 


\section*{Acknowledgements}
\indent

We thank P.~Hasenfratz, F.~Niedermayer, J.~L.~Ma\~nes and A. Labarta for
useful conversations. J.~M.~R.~is supported by a Basque Government 
F.P.I.~grant. Financial support from CICYT, contract AEN95-0590 and from 
CIRIT, contract GRQ93-1047 is also acknowledged.


\appendix


\Appendix{Primitive Translations Effects}
\indent

In this Appendix we discuss the important consequences that primitive 
translations which map points with opposite magnetisations have in the 
effective lagrangian for antiferromagnets. Let us call $\tau$ to one such 
translations. According to the procedure in section 2, $\tau$ may be 
implemented by
\be
   \ba{rccl}
   \tau : & U(x)  &  \rightarrow  &  U(x)C\hd_{\tau}, 
   \ea
\ee
where $h_{\tau}$ is a compensating $U(1)$ element that keeps the
transformed field in the coset. This implies
\be
   \ba{rl}
   \tau : &
      \left\{ \ba{ccc}
      a_{\mu}^{-} & \rightarrow & -a_{\mu}^{+}  \\
      a_{\mu}^{3}  & \rightarrow & -a_{\mu}^{3}.
      \ea \right.
   \ea
\ee
This transformation differs from $T$ in (\ref{a_time_transf}) only in terms 
with time derivatives. If we consider $\tau T$ instead of $\tau$ as an 
independent generator, we have
\be
   \ba{rl}
   \tau T: & 
      \left\{ \ba{ccc}
      a_{\mu}^{-} & \rightarrow & a_{t\mu}^{-}  \\
      a_{\mu}^{3} & \rightarrow & a_{t\mu}^{3}.
      \ea \right.
   \ea
\label{a_AF_tauT_transf}
\ee
Consequently, terms with an odd number of time derivatives are forbidden.  
Moreover, it is worth mentioning that the remaining terms for the 
antiferromagnetic effective lagrangian are the same as those 
for the ferromagnetic one.
This statement holds from considering $\t\xi$,
\be
   \ba{rl}
    \tau \xi: & 
      \left\{ \ba{ccc}
      a_{\mu}^{-} & \rightarrow & a_{\xi\mu}^{-}  \\
      a_{\mu}^{3} & \rightarrow & a_{\xi\mu}^{3},
      \ea \right.
   \ea
\label{a_AF_tauxi_transf}
\ee
where $\xi$ stands for elements of the  point group which map points with 
opposite magnetisations, like that in (\ref{a_AF_space_transf}b), together 
with $T$ and $\t T$ as the symmetry generators.

The transformations $\t\xi$ and $\t T$, given by (\ref{a_AF_tauxi_transf}) 
and (\ref{a_AF_tauT_transf}) respectively, where proposed in \cite{Burgess} 
as the macroscopic transformation for the antiferromagnetic spin waves. 
Notice that these transformations are slightly less restrictive than the ones  
we have, namely, $\t\xi$, $T$ and $\t T$.

The introduction of the spin-orbit interaction does not invalidate the 
previous statements. Indeed, as mention before the differences between $T$ 
and $\t$ come from the time indices. However, spin-orbit sources have no time 
indices and, therefore, they transform trivially under $\t T$.

Nevertheless, the statements above must be generalised in the presence of 
electromagnetic fields. The electromagnetic fields transform trivially under 
$\t$, and therefore, under $\t T$ the magnetic field, ${\bf B}$, changes  
sign. This implies that the invariants with an odd number of time 
derivatives plus magnetic fields are not allowed in the effective lagrangian 
for the antiferromagnetic spin waves. The remaining terms in the effective 
lagrangian are the same as the ones for the ferromagnetic spin waves.  


\Appendix{Mathematical Properties}
\indent

In this appendix we discuss a few important technicalities which have been
used through the paper.

The projectors $P_{\pm}$ are introduced in sect.~2 in order to
single out the magnetisation direction. They are given by
\be
P_+ =
   \left( \ba{cccc}
   1 &   &        &   \\
     & 0 &        &   \\
     &   & \ddots &   \\
     &   &        & 0
   \ea \right)
\qquad , \qquad
P_- =
   \left( \ba{cccc}
   0 &   &        &   \\
     & 0 &        &   \\
     &   & \ddots &   \\
     &   &        & 1
   \ea \right),
\ee
and verify
\be
tr(P_{\pm} S^a) = \pm s\d^{a3}.
\ee

When two of these projectors are in a trace, we can split it in two
pieces
\be
tr \Bigl( P_{\pm} (\cdots) P_{\pm} (\cdots) \Bigr) =
tr \Bigl( P_{\pm} (\cdots) \Bigr)
tr \Bigl( P_{\pm} (\cdots) \Bigr).
\ee

This is immediate because the only non-zero element in the matrix
$P_+(\cdots)P_+$ is the $(\cdots)_{11}$, or $(\cdots)_{(2s+1)\,(2s+1)}$
for the $P_-$ case.

Another important property is that
in any product of generators between two
$P_+$ where at most two of them are different from $S^3$ the
following substitution can be performed:
\bea
S^{\a} & \rightarrow & {\sqrt{2s} \over 2}
      \left( \s^{\a} \oplus 0 \right) \qquad  \qquad  (\a=1,2)
                                                           \nonumber \\
S^3    & \rightarrow & (s-1) \left({\bf 1} \oplus 0 \right) + P_+,
\eea
since in this case only the upper-left $2 \times 2$ matrix contributes
to the trace. An analogous property holds for $P_-$.

In addition in section 2 the matrix $C = e^{-i\pi S^2}$ is introduced to
implement the space and time inversions. Below we
give some relevant properties of this matrix.

The time inversion implies \cite{Pascual}
\be
C^{\da} S^a C = -(S^a)^T.
\ee

The action of $C$ over the projectors is given by
\be
C^{\da} P_+ C = P_-,
\ee
which follows from the fact that $P_{\pm}$ belong to the subspace
spanned by arbitrary
powers of $S^3$. We
also have that
\bea
tr (P_+ S^{a_1} \cdots S^{a_n} P_+) & = &
(-)^n tr (P_- S^{a_n} \cdots S^{a_1} P_-),
\eea
which follows immediately by inserting $C^{\da}C$ between the elements
on the l.h.s.~and using the properties above.

By using this property together with the clausure relations for the
Pauli matrices,
\be
{1 \over 2} (\s_+)_{\a\beta}(\s_-)_{\g\d} +
{1 \over 2} (\s_-)_{\a\beta}(\s_+)_{\g\d} =
- \d_{\a\beta}\d_{\g\d} - \s^3_{\a\beta}\s^3_{\g\d} +
                                        2\d_{\a\d}\d_{\g\beta},
\label{clausure_relation}
\ee
the relations $F_{\m\n} \sim \apm a_{\n}^- - a_{\n}^+ \amm$ and
$D_{\m}a_{\n}^- = D_{\n} \amm$ mentioned in section 2 can be proved from
the explicit representation (\ref{a_explicit}).


\Appendix{Fundamental Representation}
\indent

In this appendix we point out the simplifications that
occur for $s=1/2$ which follow from the properties of the Appendix B.

All the invariants can be constructed in terms of
\be
T(x) = U(x) \s^3 U^{\da}(x),
\ee
which transforms covariantly under $SU(2)$, their derivatives and
topological terms, which are built out of $\bm$ alone. Notice that
\be
\bm = tr (\UdidU \s^3).
\ee

Recall that $T$ verifies
\be
T^2 = 1
\qquad , \qquad
tr T = 0.
\ee

In the fundamental representation $P_+ = ({\bf 1} + \s^3)/2$ and the
building blocks are:
\be
\amm(x) =  {1 \over 4} tr
\left( [\UdidU,\s_-]\s^3 \right) =
-{i \over 4} tr (U^{\da} \pa_{\m}T U \s_-).
\label{amm}
\ee

Let us consider the covariant derivative
$D_{\n} = \pa_{\n} - i a^3_{\n}$ on terms of the form (\ref{amm}),
\be
tr (U^{\da}AU \s_-)
\qquad , \qquad
A \rightarrow gAg^{\da} \in {\cal L}(SU(2)),
\ee
where $A$ transforms covariantly under $SU(2)$ and is in the Lie
algebra. The result of the covariant derivative is
\be
D_{\n} tr (U^{\da}AU \s_-) =
tr (U^{\da}\pa_{\n}AU\s_-)
- {1 \over 2} tr (AT) tr (U^{\da}\pa_{\n}TU\s_-).
\label{result_B}
\ee

Since all the terms can be derived from (\ref{amm}) the most general
form of A is
\be
A = \pa_{\m_1}\cdots\pa_{\m_n}T,
\ee
which leads us to write for the $T$s
\bea
D_{\n} tr (U^{\da}\pa_{\m_1}\cdots\pa_{\m_n}T U \s_-) & = &
tr (U^{\da}\pa_{\n}\pa_{\m_1}\cdots\pa_{\m_n}T U\s_-)  \nonumber \\
       & & \mbox{} - {1 \over 2} tr (\pa_{\m_1}\cdots\pa_{\m_n}T T)
                      tr (U^{\da}\pa_{\n}TU\s_-).
\eea

In order to get $SU(2)$ invariants we will have products of the form
\bea
\lefteqn{ tr (U^{\da}AU \s_+) tr (U^{\da}BU \s_-) = } \qquad\quad \nn \\
& & \phantom{ + }{\ \,}
    {1 \over 2} \left[ tr (U^{\da}AU \s_+) tr (U^{\da}BU \s_-) +
          tr (U^{\da}AU \s_-) tr (U^{\da}BU \s_+) \right]    \\
& & \mbox{}
    + {1 \over 2} \left[ tr (U^{\da}AU \s_+) tr (U^{\da}BU \s_-) -
      tr (U^{\da}AU \s_-) tr (U^{\da}BU \s_+) \right], \nonumber
\eea
which have been written as the sum of its symmetric and
antisymmetric parts

Now the Pauli matrices $\s_{\pm}$ can be eliminated with the aid of the
clausure relation (\ref{clausure_relation}). For the symmetric part it
is immediate, leading to
\be
- tr(AT) tr(BT) + 2tr(AB),
\ee
while for the antisymmetric part a little trick has to be used in order
to be able to use the clausure relation: we perform in the first factor
of each term the substitution $[\s^3,\s_{\pm}] = \pm 2 \s_{\pm}$, and we
obtain
\be
\left[ tr (U^{\da}AU [\s^3,\s_+]) tr (U^{\da}BU \s_-) +
        tr (U^{\da}AU [\s^3,\s_-]) tr (U^{\da}BU \s_+) \right] =
-4 tr ([A,B] T).
\ee

Hence we conclude that for the $s = 1/2$ case all the invariants can be
constructed from products of the following traces:
\be
   \ba{lr}
   tr \left( \pa_{\m_1}\cdots\pa_{\m_n}T
                       \pa_{\n_1}\cdots\pa_{\n_m}T \right) \\
   tr \left( [\pa_{\m_1}\cdots\pa_{\m_n}T,
                       \pa_{\n_1}\cdots\pa_{\n_m}T] T \right)
   \ea
\qquad \quad
(n,m = 0,1,2,\ldots),
\ee
which correspond to take into account the symmetric and antisymmetric
parts of the product $\pa_{\m_1}\cdots\pa_{\m_n}T
\pa_{\n_1}\cdots\pa_{\n_m}T$ respectively.


\Appendix{Equivalence with the $O(3)$-sigma model}
\indent

Let us next make contact with the so called $O(3)$ $\sigma$-model formulation, 
which uses a unitary vector $n^a(x)$ as the basic building block of the
effective lagrangian. Recall first that a formula like (\ref{amm}) exist 
for any representation,
\be
\amm(x) ={1 \over 2s}  tr
\left( [\UdidU,S_-]P_{+} \right) =
-{i \over 2s}  tr (U^{\da} \pa_{\m}T_{+} U S_{-})
\quad , \quad
T_{+}=UP_{+}U^{\dagger}.
\label{amm_C}
\ee

We can relate our notation with the standard one of the unitary vector
$n^a(x)$ by noting that
\be
T(x) = U(x) S^3 U^{\da}(x) = n^a(x) S^a.
\label{T_definition}
\ee

It is easy to check that $\nv^2=1$ and clearly $n^a$ transforms like
a vector. Recall next that $P_{+}$ can be written as
\be
P_{+}=\sum_{n=0}^{2s} a_{n}(S^3)^{n}
\ee
since $\{ (S^3)^{n}\} $ is a bases of the subspace of diagonal matrices.
Hence
\be
T_{+}=\sum_{n=0}^{2s} a_{n}(US^3U^{\dagger})^{n}=
\sum_{n=0}^{2s} a_{n}T^n.
\ee

For an arbitrary representation the covariant derivative
(\ref{a_covariant_derivative})
on a term like (\ref{amm_C}) yields a formula similar to that in
(\ref{result_B}),
\be
D_{\n} tr (U^{\da}AU S_-) = tr (U^{\da}\pa_{\n}AU S_-)
           - {1 \over s} tr (AT) tr (U^{\da}\pa_{\n}T_+ US_-),
\ee
where $A$ is made of an arbitrary number of derivatives on $T_+$. The
possible invariants will be made out of products of the form
\bea
\lefteqn{ tr (U^{\da}AU S_+) tr (U^{\da}BU S_-) = } \qquad \quad
                                                    \nonumber \\
 & & \phantom{ + } {\ \,}
     {1 \over 2} \left[ tr (U^{\da}AU S_+) tr (U^{\da}BU S_-) +
           tr (U^{\da}AU S_-) tr (U^{\da}BU S_+) \right]
                                              \label{product} \\
 & & \mbox{}
     + {1 \over 2} \left[ tr (U^{\da}AU S_+) tr (U^{\da}BU S_-) -
       tr (U^{\da}AU S_-) tr (U^{\da}BU S_+) \right]. \nonumber
\eea

Taking into account that $U \in SU(2)$ the property below follows:
\be
(US^aU^{\da})(US^aU^{\da})=(S^a)(S^a),
\label{S^2(1)}
\ee
which can be rewritten as
\bea
\lefteqn{
{1 \over 2} \left[ (US_+U^{\da})(US_-U^{\da}) +
    (US_-U^{\da})(US_+U^{\da}) \right] = -(T)(T) }
\qquad \qquad \qquad \qquad \quad     \label{S^2(2)} \\
 & & \mbox{} + {1 \over 2}\Bigl[ (S_+)(S_-) + (S_-)(S_+) \Bigr] +
                                          (S^3)(S^3).  \nonumber
\eea

Using (\ref{S^2(2)}) the symmetric part of (\ref{product}) is
\be
-tr(AT)tr(BT) + {1 \over 2} \Bigl[ tr(AS_+)tr(BS_-) +
              tr(AS_-)tr(BS_+) \Bigr] + tr(AS^3)tr(BS^3),
\ee
whereas for the antisymmetric part the same trick as in Appendix C has
to be used, i.e., change $S_{\pm} = \pm [S^3,S_{\pm}]$ in the first
factor of each term. We obtain
\bea
\lefteqn{{1 \over 2} \left[ tr (U^{\da}AU [S^3,S_+]) tr (U^{\da}BU S_-)
  + tr (U^{\da}AU [S^3,S_-]) tr (U^{\da}BU S_+) \right] = }
\qquad \quad  \\
 & & {1 \over 2} \Bigl[ tr([A,T]S_+)tr(BS_-) +
       tr([A,T]S_-)tr(BS_+) \Bigr] + tr([A,T]S^3)tr(BS^3).  \nonumber
\eea

Finally it has been shown that all the invariants can be written in
terms of $T$ and therefore in terms of $n^a(x)$.
The reverse is also true. Suppose we have written our
effective lagrangian in terms of derivatives of $n^a$. From
(\ref{T_definition})
\be
n^a = {3 \over s(s+1)(2s+1)} tr(S^3 U^{\dagger}S^a U).
\ee

Consider a vector of the form
\be
v^a = tr(A U^{\dagger}S^a U)
\qquad , \qquad
A\in {\cal L}(SU(2)),
\label{vector}
\ee
\be
\partial_{\m} v^a = tr(\partial_{\mu} A U^{\dagger}S^a U)
+ tr([ U^{\da} \pa_{\m} U , A ] U^{\dagger}S^a  U).
\ee

Since any vector will be obtained by applying several derivatives on $n^a$, 
$A$ in (\ref{vector}) will only contain $ U^{\da} \pa_{\m} U $ and its
derivatives, which can be written in terms of our basic fields 
$a_{\m}^{\pm}, a_{\m}^3$. The remaining dependence on $U$ is through 
$U^{\dagger}S^a U$ only which will cancel out upon contraction with other 
vectors, according to (\ref{S^2(1)}), in order to build a scalar lagrangian.

So far we have been talking about invariant terms in the effective lagrangian. 
It is not true however that terms which are invariant up to a total
derivative, like $a_{\m}^3$, can be written in terms of $n^a$ in a local form.
However, we can write them locally in $T_{+}$ or $n^a$ if we introduce an 
extra dimension in the following way. We interpolate smoothly the Goldstone 
fields $\pi^{\alpha}(x)\rightarrow \pi^{\alpha}(x,\lambda)$, $\l \in [0,1]$
in such a way that $\pi^{\alpha}(x,1)=\pi^{\alpha}(x)$ and 
$\pi^{\alpha}(x,0)=0$. Let us concentrate on $ a_{0}^3 $ which is the only 
one which arises in our effective lagrangian. It is very easy to check that
\be
\bo = {1 \over s} \int_0^1 d\lambda \epsilon^{\alpha\beta}
            tr(T_{+} \partial_{\alpha}T_{+} \partial_{\beta}T_{+}) \sim
  \int_0^1d\lambda \epsilon^{\alpha\beta}
      \epsilon_{abc} n^{a} \partial_{\alpha}n^{b} \partial_{\beta}n^{c},
\label{interpolation}
\ee
$(\alpha, \beta = 0, \lambda)$. The second  equality follows upon using 
(\ref{T_definition}) and performing the trace. The final result must be a 
scalar function of $n^{a}$, $\partial_{\a}n^{b}$ and $\partial_{\beta}n^{c}$ 
antisymmetric under the exchange of $\alpha$ and $\beta$, being 
(\ref{interpolation}) the only possibility. To our knowledge this term was 
first written in the last form in \cite{Volovik} (see also \cite{LeutwylerNR}),
 whereas we have not been able to locate the two previous forms in the 
literature. We stress again that only the form that we use in our effective 
lagrangian is local. It is usual however to find (\ref{interpolation}) with a 
rather different aspect, namely,
\be
\bo \sim A^a (n)\partial_0 n^a
\qquad , \qquad
n^{a}=\epsilon_{abc}{\partial A^{b}\over \partial n^{c}},
\label{n_implicit}
\ee
where the second equation gives $A^a$ as an implicit function of $n^b$ 
\cite{Moon,Fradkin}. It is easy to check that the last expression of 
(\ref{interpolation}) and (\ref{n_implicit}) give rise to the same equations 
of motion.



\Appendix{Coupling to Constant Electric and Mag\-netic Fields}
\indent

In this Appendix we present a further example on how the effective lagrangian
for spin waves coupled to electromagnetism can be used to obtain qualitative
information on the system. 

Let us suppose that the system is exposed to constant electric and magnetic 
fields and address the question on how the low momentum dispersion relations 
of the spin waves change in the ferromagnetic and antiferromagnetic case. For 
simplicity we assume that the magnetic field is on the third direction and
the electric field in the $z-\bar z$ plane, and fix their relative size, which
is now arbitrary, as $eaE/J \sim ea^2B $. Recall that the Pauli term must be 
counted as $\mu B/J \sim 10 ea^2B$. We also neglect terms induced by 
spin-orbit or magnetic dipole interactions which explicitly break the 
$SU(2)$ symmetry since it is straightforward to take them into account if
desired. We shall focus in the leading effects due to higher order terms, 
since those due to the lowest order terms are well known and will be easily 
reproduced.

For the ferromagnetic case the leading corrections due to higher order terms 
arise from
\bea 
& & \ba{l}
    (\apz \amzb + \amz \apzb) E^z E^{\zb} \\
    \ap3 \am3  E^z E^{\zb}  \\ 
    \apz \amz E^z E^z + \amzb \apzb E^{\zb} E^{\zb}   \\
    (\apz \am3 + \amz \ap3) E^{\zb} E^{\zb} + (\amzb \ap3 + \apzb \am3) E^z E^z \\
    (\apz \amzb + \amz \apzb) B^3 B^3  \\
    \ap3 \am3  B^3 B^3,
    \ea 
\eea
which together with the contributions of the leading order terms 
(\ref{space_a2}) and (\ref{time_a1}) lead to
\bea
\o & = & \m B^3 + {2 \over m} [1 + \e_1 E^z E^{\zb} 
                                 + \beta_1 B^3 B^3] k^z k^{\zb} \nn \\
& & \phantom{\m B^3} + {1 \over 2\g m} [1 + \e_2 E^z E^{\zb} 
                                     + \beta_2 B^3 B^3] (k^3)^2 \nn \\
& & \phantom{\m B^3} + \e_3 [(E^z k^{\zb})^2 + (E^{\zb} k^z)^2]   
                                                  \label{dr_fields} \\
& & \phantom{\m B^3} + 2 \e_4 [(E^z E^z k^z) + (E^{\zb} E^{\zb} k^{\zb})] k^3. 
                                                                \nn 
\eea

The effect of our particular constant electric and magnetic fields is, apart 
from creating the well known energy gap proportional to the magnetic field,
(i) renormalising $1/m$ and $\gamma$ and (ii) producing producing further 
anisotropies in the dispersion relation (due to $\e_3$ and $\e_4$), which 
vanish if the electric field vanishes.

For the antiferromagnetic case the leading corrections due to higher order 
terms arise from the same terms as in the ferromagnetic case (\ref{dr_fields}) 
plus the following extra terms, 
\bea 
& & \ba{l}
    \apo \amo  E^z E^{\zb} \\
    \apo \amo  B^3 B^3     \\
    i[(\apo \amz + \amo \apz) E^z  B^3 - (\amo \apzb +\apo \amzb) E^{\zb} B^3, 
    \ea 
\eea
which together with the contributions of the leading order terms 
(\ref{space_a2_AF}) and (\ref{time_a2}) lead to
\bea
\o_{\pm} & =  & \pm \m B^3  \pm i \eta[(E^z k^{\zb}) - (E^{\zb} k^z)]B^3 \nn \\
& & \phantom{ \pm \m B^3 } 
       + \Big[ 4v^2[1 + \e_1 E^z E^{\zb} + \beta_1 B^3 B^3] k^z k^{\zb} \nn \\
& & \phantom{ \pm \m B^3 + \Big[ }                    
               + (\g v)^2  [1 + \e_2 E^z E^{\zb} + \beta_2 B^3 B^3] (k^3)^2 \\
& & \phantom{ \pm \m B^3 + \Big[ }
               + \e_3 [(E^z k^{\zb})^2 + (E^{\zb} k^z)^2]  \nn \\
& & \phantom{ \pm \m B^3 + \Big[ }
               + 2 \e_4 [(E^z E^z k^z) 
                         + (E^{\zb} E^{\zb} k^{\zb})] k^3 \Big]^{1/2}. \nn
\eea

The effect of our particular constant electric and magnetic fields is, apart 
from providing the well known "chemical potential" term proportional to the 
magnetic field, (i) renormalising $v$ and $\gamma$, (ii) producing further 
anisotropies in the dispersion relation (due to $\e_3$ and $\e_4$), which 
vanish if the electric field vanishes, and (iii) producing a momentum 
dependent distinction between the two components of the spin wave (due to 
$\eta$). We find the latter effect particularly interesting. Notice that it 
requires both the electric and magnetic fields be differnt from zero.




\end{document}